\documentclass[preprintnumbers,amsmath,amssymb,prd,superscriptaddress,nofootinbib,longbibliography]{revtex4}
\usepackage{color}
\usepackage{epsfig}
\usepackage{graphicx}
\DeclareGraphicsExtensions{.eps,.ps,.eps.gz,.ps.gz,.eps.Z,{}}

\usepackage{dcolumn}
\usepackage{bm}

\newcommand{\vx}{{\bf x}}

\newcommand{\vd}{{\bf d}}

\newcommand{\vq}{{\bf q}}

\newcommand{\vu}{{\bf u}}

\newcommand{\vk}{{\bf k}}
\newcommand{\ii}{{\rm i}}
\newcommand{\dd}{{\rm d}}

\newcommand{\mg}{\big<}
\newcommand{\md}{\big>}

\newcommand{\mB}{{\cal B}}

\newcommand{\mF}{{\cal F}}
\newcommand{\mH}{{\cal H}}

\newcommand{\Dirac}{\delta_{\rm D}}

\newcommand{\beq}{\begin{equation}}
\newcommand{\eeq}{\end{equation}}
\newcommand{\beqa}{\begin{eqnarray}}
\newcommand{\eeqa}{\end{eqnarray}}

\def\fun#1#2{\lower3.6pt\vbox{\baselineskip0pt\lineskip.9pt
        \ialign{$\mathsurround=0pt#1\hfill##\hfil$\crcr#2\crcr\sim\crcr}}}

\def\Mpc{\, h^{-1} \, {\rm Mpc}}

\newcommand{\bdm}{\begin{displaymath}}
\newcommand{\edm}{\end{displaymath}}
\newcommand{\bea}{\begin{eqnarray}}
\newcommand{\eea}{\end{eqnarray}}
\newcommand{\bt}{\begin{tabular}}
\newcommand{\et}{\end{tabular}}

\newcommand{\oneloop}{{\rm 1-loop}}
\newcommand{\twoloops}{{\rm 2-loop}}
\newcommand{\ct}{{\rm c.t.}}

\newcommand{\reg}{{\rm reg}}
\newcommand{\tree}{{\rm tree}}
\newcommand{\sym}{{\rm sym.}}
\newcommand{\init}{{\rm init.}}
\newcommand{\adiab}{{\rm adiab.}}
\newcommand{\dpartial}{{\delta}}
\newcommand{\lbox}{{\rm box}}
\newcommand{\bin}{{\rm bin}}

\newcommand{\sigmav}{{\sigma_{\rm displ.}}}
\newcommand{\sigmadv}{{\sigma^{2}_{\rm displ.}}}
\newcommand{\etaf}{{\eta_{f}}}
\newcommand{\etai}{{\eta_{i}}}

\begin{document}
\title{Constructing Regularized Cosmic Propagators}
\author{Francis Bernardeau}
\affiliation{Institut de Physique Th\'eorique,
 CEA, IPhT, F-91191 Gif-sur-Yvette, France
 CNRS, URA 2306, F-91191 Gif-sur-Yvette, France}
\author{Mart\'{\i}n Crocce}
\affiliation{Institut de Ci\`encies de l'Espai, IEEC-CSIC, Campus UAB,
Facultat de Ci\`encies, Torre C5 par-2,  Barcelona 08193, Spain}
\author{Rom\'an Scoccimarro}
\affiliation{Center for Cosmology and Particle Physics,  \\
Department of Physics, New York University, New York, NY 10003}

\vspace{.2 cm}
\date{\today}
\vspace{.2 cm}
\begin{abstract}

We present a new scheme for the general computation of cosmic
propagators that allow to interpolate between standard  perturbative
results at low-$k$ and their expected large-$k$ resummed
behavior. This scheme is applicable to any multi-point propagator  and
allows the matching of perturbative low-$k$ calculations to any number
of loops to their large-$k$ behavior, and can potentially be applied
in case of non-standard cosmological scenarios such as those with
non-Gaussian initial conditions.  The validity of our proposal is
checked against previous prescriptions and measurements in numerical
simulations  showing a remarkably good agreement.  Such a
generic prescription for multi-point propagators provides the
necessary building blocks for the computation of  polyspectra in the
context of the so-called $\Gamma$-expansion  introduced by
 Bernardeau et al. (2008). As a  concrete application
we present a consistent calculation of the matter bispectrum at one-loop order. 

\end{abstract}
 \vskip2pc

\maketitle

\section{Introduction}

The large-scale structure of the universe that we observe today is thought to emerge from gravitational instabilities out
of primordial metric perturbations, therefore precise observations of
the large-scale structure of the local universe can be used to put
constraints on  cosmological models. The connection between
cosmological parameters and observations are however only
straightforward when they correspond to the linear regime, i.e. when
the observables can be computed as a linear transform of the
primordial metric perturbations; this connection is less
trivial when nonlinearities, due in a large part to the gravitational dynamics itself, are present. Those nonlinearities arise  during the late stage of the gravitational instabilities. This is an epoch during which the universe can be safely assumed to be matter dominated (at least in the context of the standard model of cosmology where the dark energy  component forms an homogeneous fluid.)

According to the cosmological principle  cosmic fields are statically homogeneous and isotropic. In the context we are interested in, two degrees of freedom are then relevant (see~\cite{2002PhR...367....1B} for details): the local density contrast, $\delta(\vx)$ and the velocity divergence, $\theta(\vx)=\partial_{i}\vu_{i}(\vx)$.  It is then very fruitful to introduce the Fourier modes of the cosmic fields $\delta(\vk)$ and $\theta(\vk)$, which evolve independently of one another in the linear regime. One can then introduce the doublet 
$\Psi_{a}(\vk)=\left\{\delta(\vk),-\theta(\vk)/\mH\right\}$, where
${\cal H}$ is the conformal expansion rate with its time evolution described by the Friedmann equation, to write down the equations of motion in compact form and facilitate the implementation of resummation techniques. 

In this context, the goal of the theoretical and numerical calculations is a precise description, beyond the linear regime, of the statistical properties of $\Psi_{a}$. We are  particularly interested in the multi-component power  spectra $P_{ab}(k)$ defined as,
\begin{equation}
\mg \Psi_{a}(\vk)\Psi_{b}(\vk')\md=\Dirac(\vk+\vk')P_{ab}(k)
\end{equation}
where the Latin indices $a$ and $b$ vary from 1 to 2, which are implicitly or explicitly measured in observations such as galaxy surveys, and 
 higher-order spectra such as bispectra,
\begin{equation}
\mg \Psi_{a}(\vk_{1})\Psi_{b}(\vk_{2})\Psi_{c}(\vk_{3})\md=\Dirac(\vk_{1}+\vk_{2}+\vk_{3})B_{abc}(\vk_{1},\vk_{2}).
\end{equation}

This problem is in general very complicated if one wants to solve it from first principles. It can be made slightly easier to
address in the context of the standard cosmological model where the primordial metric perturbations are expected to follow Gaussian or near Gaussian statistics. In that case the primordial properties of the fields are entirely determined by the initial power spectra, $P_{ab}^{\init}(k)$.
The question is then to uncover the functional dependence of $P_{ab}(k)$ as time evolves with $P_{ab}^{\init}(k)$.

In the last few years attempts have been made to present perturbative schemes in the context of the growth of structure.
The standard perturbation theory is unambiguously defined but leads to uncontrollable results \cite{2002PhR...367....1B}. On the other hand alternative approaches have been proposed which produce more robust results, such as the Renormalized Perturbation Theory (RPT) in
\cite{2006PhRvD..73f3519C}, the Time Renormalization Group (TRG) approach in \cite{2008JCAP...10..036P}, the closure theory in \cite{2009PhRvD..79j3526H,2008ApJ...674..617T}, or with the help of perturbation theory expansion expressed in Lagrangian variables 
\cite{2008PhRvD..77f3530M,2008PhRvD..78h3503B,2011JCAP...08..012O,2011PhRvD..83h3518M},  etc. One of such approaches, advocated in~\cite{2008PhRvD..78j3521B}, is based on what has been called the $\Gamma$-expansion.  This method 
exploits the following relation,
\begin{equation}
\mg \Psi_{a}(\vk)\Psi_{b}(\vk')\md
=\sum_{p}\frac{1}{p!}
\int\dd^3\vq_{1}\dots \dd^3\vq_{p}
\left\langle \frac{\dpartial \Psi_{a}(\vk)}{\dpartial\Phi_{a_{1}}(\vq_{1})\dots \dpartial\Phi_{a_{p}}(\vq_{p})}\right\rangle
P^{\init}_{a_{1}b_{1}}(q_{1})\dots P^{\init}_{a_{p}b_{p}}(q_{p})
\left\langle \frac{\dpartial \Psi_{b}(\vk')}{\dpartial\Phi_{b_{1}}(\vq_{1})\dots \dpartial\Phi_{b_{p}}(\vq_{p})}\right\rangle
\label{PkRecons}
\end{equation}
where the standard Einstein convention (repeated indices are summed over) is used and 
where $\Phi_{a}(k)$ is the value of $\Psi_{a}(k)$ at the initial time. This relation 
expresses the fact that ensemble average over primordial fluctuations
can be reorganized in an alternative way to that in standard
perturbation theory \cite{2002PhR...367....1B}.
It exhibits the multi-point propagators defined as the ensemble average of the 
infinitesimal variation of the cosmic fields 
with respect to the initial conditions. More precisely  the multi-point propagators\footnote{It is important to note that in this paper we call
the $\Gamma^{(p)}$ functions the $(p+1)$-propagators -- because it connects $p+1$ lines -- but alternative conventions can be found in the literature where $\Gamma^{(p)}$
is called the $p$-point propagator.} $\Gamma^{(p)}(\vq_{1},\dots,\vq_{p})$ are defined as
\begin{equation}
\frac{1}{p!}\left\langle \frac{\dpartial \Psi_{a}(\vk)}{\dpartial\Phi_{a_{1}}(\vq_{1})\dots \dpartial\Phi_{a_{p}}(\vq_{p})}\right\rangle
\equiv \Dirac\Big(\vk-\sum_{r=1}^{p}\vq_{r}\Big)\, \Gamma^{(p)}(\vq_{1},\dots,\vq_{p}).
\end{equation}
The relation (\ref{PkRecons}) is valid for Gaussian initial conditions
but can be extended  for non-Gaussian initial conditions~\cite{2010PhRvD..82h3507B}. It clearly shows that
the propagators are key ingredients for calculating nonlinear power spectra. They are the building blocks of the $\Gamma-$expansion approach~\cite{2008PhRvD..78j3521B} and the focus of this paper. 

The second reason why these quantities appear to be important ingredients in perturbation theory calculations is that their 
asymptotic properties, e.g. how they behave for large wave-numbers, can be computed beyond perturbation expansions. This has been pioneered
in \cite{2006PhRvD..73f3519C,2006PhRvD..73f3520C} and recently
 revised in \cite{2011arXiv1109.3400B} with the so-called eikonal approximation. More specifically it has been shown that in the high-$k$ limit the multi-point propagators are damped by a function that depends on the displacement field alone, irrespectively 
of the dynamics responsible of this displacement. Predictions on the behavior of those objects are then robust and can be computed in various approximations.

This motivates their use as the building blocks of a perturbation theory scheme. To achieve this end, one must have a description of multi-point propagators at all scales, matching the perturbative calculations at low-$k$ to the resummed asymptotic behavior at high-$k$.  As resummed propagators also contain information on loop contributions, both regimes can only be matched if a consistent interpolation schemes can be built. Furthermore, it is arguably this interpolation regime the most important for the prediction of loop corrections to (equal-time) correlation functions, as e.g. the power spectrum: while the low-$k$ limit can be safely computed using standard perturbation theory, the high-$k$ limit only adds a (time-dependent) phase-shift to Fourier modes and thus does not contribute to  the power spectrum or bispectrum. 

This problem  has been solved for the two-point propagator in \cite{2006PhRvD..73f3520C} with the help of an exponentiation scheme that interpolates between the one-loop
results and its high-$k$ behavior.  However, this  prescription is somewhat ad-hoc in that is specific  to matching one-loop to resummed behavior of the two-point 
propagator  but it is not clear a priori how it can be extended  to incorporate  higher-loop information at low-$k$ or its generalization to multi-point propagators. The aim of this paper is to revisit this problem and propose a consistent solution which would be valid for any propagators
and incorporate perturbative information to any loop order.

This paper is organized as follows. In the section II we recall the
basic  equations of motion, define our notation including the
diagrammatic description and present the $\Gamma$-expansion of power
spectra and bispectra in terms of multi-point propagators.  In section
III we review the expected properties of the propagators. In section
IV the proposed interpolation scheme is presented in detail,
 while a comparison of our theoretical predictions for the
tree-point propagator to numerical simulations is presented in section V.
The implications of our results are illustrated with a bispectrum
computation in the  section VI. Lastly section VII contains our conclusions.

\section{Equations of motion and the $\Gamma$-Expansion} 

\subsection{The equations of motion}

We are interested here in the early stages of the development of cosmological instabilities in a cosmological dust fluid. In general the dynamical evolution of such a fluid can be described with the Vlasov equation for which one further assumes that multi-flow regions play a negligible role (see e.g.~\cite{PueSco0908,2009arXiv0910.1002M,2010arXiv1004.2488B,PieManSav1108} for recent discussion on going beyond this).  In the single flow limit, the  equations of motion then takes the
form of a set of three coupled equations relating the local density contrast, the peculiar
velocity field and the gravitational potential (see \cite{2002PhR...367....1B}). 

At {\it linear order} these equations can easily be solved for an arbitrary background cosmology.
One generically finds a growing solution and a decaying solution. Let us denote $D_{+}(\tau)$ the growing 
mode solution for the density contrast, with $\tau$   the conformal time, and $f_{+}(\tau)$ its logarithmic derivative with respect to the scale factor so that,
\begin{equation}
\delta(\vk,\eta)=D_{+}(\eta)\delta_{0}(\vk),\ \ \theta(\vk,\eta)/\mH=-f_{+}(\eta)D_{+}(\eta)\delta_{0}(\vk)
\end{equation}
is the solution for the growing mode and similarly,
\begin{equation}
\delta(\vk,\eta)=D_{-}(\eta)\delta_{0}(\vk),\ \ \theta(\vk,\eta)/\mH=-f_{-}(\eta)D_{-}(\eta)\delta_{0}(\vk)
\end{equation}
for the decaying mode.

Following \cite{1998MNRAS.299.1097S}, the  equations of motion describing a pressureless fluid in the one-flow limit can be written
in a compact form with the use of the two component quantity $\Psi_a(\vk,\tau)$, defined as 
\begin{equation}
\Psi_a(\vk,\tau) \equiv \Big\{ \delta(\vk,\tau),\ -\frac{1}{f_{+}(\tau)\cal H}\theta(\vk,\tau) \Big\},
\label{2vector}
\end{equation}
where
${\cal H}\equiv {d\ln a
/{d\tau}}$  is the conformal expansion rate  with  $a(\tau)$ the cosmological scale factor and where the 
index $a=1,2$ selects the density or velocity components. Note that this definition of $\Psi_a$ is slightly different than the one used in the introduction
and makes explicit use of the growing solution. The function $f_{+}$ is unity for an Einstein-de Sitter background only. At this stage one can also remark that the choice of this basis is somehow arbitrary: we could have use any independent linear combinations of $\delta(\vk,\tau)$
and $-\theta(\vk,\tau)/{f_{+}(\tau)\cal H}$ are our choice of doublet fields. 

It is then convenient to rewrite  the time dependence in terms of the growing solution and in the following we will use the time variable $\eta$
defined as 
\begin{equation}
\eta=\log {D_{+}(\tau)}
\end{equation}
assuming the growth factor is set to unity at initial time. Then the {\it fully nonlinear} equations of
motion in Fourier space (we henceforth use the convention that  repeated
Fourier arguments are integrated over) read \cite{2002PhR...367....1B},
\begin{eqnarray}
\frac{\partial}{\partial \eta} \Psi_a(\vk,\eta) + \Omega_{ab}(\eta) \Psi_b(\vk,\eta) &=& \gamma_{abc}(\vk,\vk_1,\vk_2) \ \Psi_b(\vk_1,\eta) \ \Psi_c(\vk_2,\eta),
\label{eom}
\end{eqnarray}
where 
\begin{equation}
\Omega_{ab} (\eta) \equiv \Bigg[ 
\begin{array}{cc}
0 & -1 \\ -\frac{3}{2}\frac{\Omega_{m}}{f_{+}^2} & \frac{3}{2}\frac{\Omega_{m}}{f_{+}^2}-1 
\end{array}        \Bigg],
\end{equation}
and the {\sl symmetrized vertex} matrix $\gamma_{abc}$ describes the non linear 
interactions between different Fourier modes. Its components are given by
\begin{eqnarray}
\gamma_{222}(\vk,\vk_1,\vk_2)&=&\Dirac(\vk-\vk_1-\vk_2) \ \frac{|\vk_1+\vk_2|^2 (\vk_1
\cdot\vk_2 )}{2 k_1^2 k_2^2}, \nonumber \\
\gamma_{121}(\vk,\vk_1,\vk_2)&=&\Dirac(\vk-\vk_1-\vk_2) \  \frac{(\vk_1+\vk_2) \cdot
\vk_1}{2 k_1^2},
\label{vertexdefinition}
\end{eqnarray}
$\gamma_{abc}(\vk,\vk_a,\vk_b)=\gamma_{acb}(\vk,\vk_b,\vk_a)$, and $\gamma=0$ 
otherwise, where $\Dirac$ denotes the Dirac delta distribution. The matrix  $\gamma_{abc}$ is independent 
on time (and on the background evolution) and encodes all the
non-linear couplings of the system.
 The formal integral solution to Eq.
(\ref{eom}) is given by (see \cite{1998MNRAS.299.1097S,2001NYASA.927...13S,2006PhRvD..73f3519C} for a detailed derivation) 
\begin{eqnarray}
\Psi_a(\vk,\eta) &=& g_{ab}(\eta) \ \Phi_b(\vk) +  \int_0^{\eta}  {\dd \eta'} \ g_{ab}(\eta,\eta') \ \gamma_{bcd}^{(\rm s)}(\vk,\vk_1,\vk_2) \Psi_c(\vk_1,\eta') \Psi_d(\vk_2,\eta'),
\label{eomi}
\end{eqnarray}
where  $\Phi_a(\vk)\equiv\Psi_a(\vk,\eta=0)$ denotes the initial conditions, set when the growth factor $D_{+}=1$ and where $g_{ab}(\eta)$ is the {\em linear propagator}, i.e. the Green's
function of the linearized version of Eq.~(\ref{eom}) and describes the standard linear evolution of
the density and velocity fields away from their initial values. 

In the following calculations we will be using the value of the $\Omega_{ab}$ matrix to be that of the Einstein de Sitter background
thus assuming that $f_{+}^{2}=\Omega_{m}$. Effectively it assumes that
$D_{-}$ scales like $D_{+}^{-3/2}$. This is known to be a very good
approximation even in the context of a $\Lambda$CDM universe. Within
this approximation $\Omega_{ab}$ becomes effectively time
independent.  It should be noticed that although the results presented below depend on this approximation, the whole construction is not based upon it. Calculations in an arbitrary background would simply make the whole presentation much more cumbersome, preventing the writing of explicit analytic forms. See Appendix A in~\cite{2008JCAP...10..036P} for this. 

The ensemble average of any quantity can then be built out of the statistical properties of the initial fields. They are 
entirely defined from the initial power spectrum of density fluctuations $P_{ab}(k)$,
\begin{equation}
\mg\Phi_{a}(\vk)\Phi_{b}(\vk')\md=\Dirac(\vk+\vk')P_{ab}(k).
\label{Spectra}
\end{equation}  
In what follows most of the calculations and applications will be made assuming initial conditions in
the growing mode, for which $\Phi_a(\vk) = \delta_0(\vk) u_a$ with $u_{a}=(1,1)$, and therefore with $P^{\init}_{ab}(k)=P_0(k) u_a u_b$.

The linear propagator $g_{ab}(\eta)$  is one of the key ingredients and gives the variation of the mode amplitude as time evolves.
The idea at the heart of the RPT approach is to generalize this operator beyond linear
theory  \cite{2006PhRvD..73f3519C,2006PhRvD..73f3520C}. More specifically the quantity ${\dpartial \Psi_{a}(\vk,\eta)}/{ \dpartial \Phi_{b}(\vk')}$
expresses the way $\Psi_{a}(\vk,\eta)$  depends on $\Phi_{b}(\vk')$ as a function of time $\eta$. This function
however depends on the stochastic properties of the fields and one is led to define its
ensemble average, $G_{ab}(\vk,\etaf,\etai)$, as
\begin{equation}
\mg\frac{\dpartial \Psi_{a}(\vk)}{\dpartial
\Phi_{b}(\vk')}\md=\Dirac{\left(\vk-\vk'\right)}\,G_{ab}(k,\etaf,\etai),
\label{Gabdef}
\end{equation} 
where we have re-introduced the initial time $\etai$. This quantity, known as the non-linear (two-point) propagator, depends on the initial fluctuations through the mode 
couplings. The ensemble 
average is made precisely over these modes. The Dirac-$\delta$
function is due, as
usual, to the homogeneity of the underlying statistical process. 

The expression for $G_{ab}$ can be computed order by order in perturbation theory.
Such results can be obtained from a formal expansion of $\Psi_{a}(\vk,\eta)$ with respect to the initial field,
\begin{equation}
\Psi_{a}(\vk,\eta)=\sum_{n=1}^{\infty}\Psi_{a}^{(n)}(\vk,\eta)
\label{PsiExpansion}
\end{equation}
with
\begin{eqnarray}
\Psi_{a}^{(n)}(\vk,\eta)=
\mF^{(n)}_{a b_1 b_2 \ldots b_n}(\vk_{1},\dots,\vk_{n};\eta)
\Phi_{b_1}(\vk_{1})\dots\Phi_{b_n}(\vk_{n})
\label{mFndef}
\end{eqnarray}
where $\mF^{(n)}$ are fully symmetric functions of the wave-vectors.  Note that these functions have
in general a non-trivial time dependence because they also include sub-leading terms in $\eta$. 
Their fastest growing term is of course given by the well known $\{F_n,G_n\}$ kernels in PT (assuming
growing mode initial conditions),
\begin{equation}
\mF^{(n)}_{a b_1 b_2 \ldots b_n}(\vk_{1},\dots,\vk_{n};\eta)u_{b_{1}}\ldots u_{b_{n}}=\Dirac(\vk-\vk_{1\dots n})\,\exp(n\eta)\ \{F_n(\vk_1,..,\vk_n),G_n(\vk_1,..,\vk_n)\} \nonumber
\end{equation}
for $a=1,2$ (density or velocity divergence fields respectively).

The concept of higher-order propagators is a natural extension of the non-linear propagator $G_{ab}$. Such
functions, that we denote $\Gamma^{(p)}_{ab_{1}\dots b_{p}}(\vk_{1},\dots,\vk_{p})$, can be defined as,
\begin{equation}
\frac{1}{2}\mg\frac{\dpartial^2 \Psi_{a}(\vk)}{\dpartial\Phi_{b}(\vk_{1})\dpartial\Phi_{c}(\vk_{2})}\md
=\Dirac(\vk-\vk_{1}-\vk_{2})\Gamma^{(2)}_{abc}\left(\vk_{1},\vk_{2}\right)
\label{GammaabcDef}
\end{equation}
for  second order (or three points), and for an arbitrary order they read,
\begin{eqnarray}
\frac{1}{p!}
\mg\frac{\dpartial^p
\Psi_{a}(\vk)}{\dpartial\Phi_{b_{1}}(\vk_{1})\dots\dpartial\Phi_{b_{p}}(\vk_{p})}\md = \Dirac(\vk-\vk_{1 \ldots p})\Gamma^{(p)}_{ab_{1}\dots
b_{p}}\left(\vk_{1},\dots,\vk_{p}\right),
\label{GammaAllDef}
\end{eqnarray}
where $\vk_{1\ldots p}=\vk_1+\ldots + \vk_p$. They can be viewed as the building blocks of the theory. Note that for the purposes we consider here, we restricted our definition to derivatives  with respect to the initial fields but a much more general description could be adopted.

It is probably worth mentioning that $\Gamma^{(p)}_{ab_{1}\dots
b_{p}}\left(\vk_{1},\dots,\vk_{p}\right)$ are real functions which, for parity reasons, obey
\begin{equation}
\Gamma^{(p)}_{ab_{1}\dots
b_{p}}\left(-\vk_{1},\dots,-\vk_{p}\right)=\Gamma^{(p)}_{ab_{1}\dots
b_{p}}\left(\vk_{1},\dots,\vk_{p}\right).\label{GammapParity}
\end{equation}

\subsection{Diagrammatic representations}

\begin{figure}
\centerline{\epsfig {figure=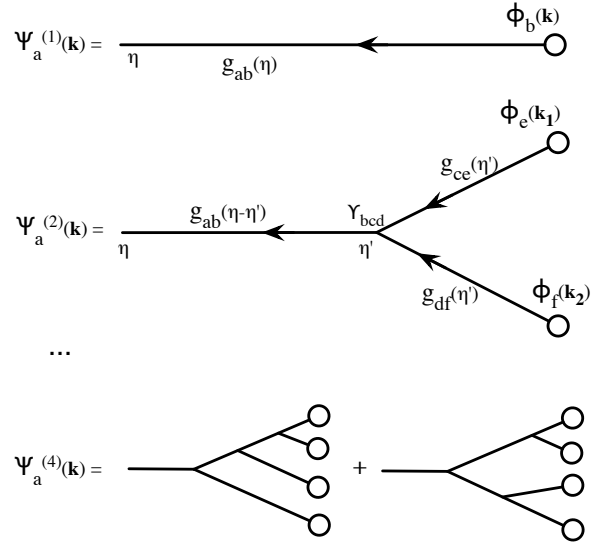,width=8cm}}
\caption{Diagrammatic representation of the series expansion of $\Psi_{a}(\vk)$ up to fourth order in the initial conditions $\Phi$. Time increases along each segment according to the arrow and each segment bears a factor $g_{cd}(\etaf-\etai)$ if $\etai$ is the initial time and $\etaf$ is the final time. 
At each initial point and each vertex point there is a sum over the component indices; a sum over the incoming wave modes is also implicit and, finally, the time coordinate of the vertex points is integrated from $\etai=0$ to the final time $\etaf$ according to the time ordering of each diagram. For instance, at fourth order there are two different possible topologies.}
\label{DiagramPsiExpansion}
\end{figure}

A detailed description of the procedure to draw diagrams and compute their values can be found in \cite{2006PhRvD..73f3519C}, we can briefly summarize these rules here as follows. In Fig. \ref{DiagramPsiExpansion} the open circles represent the initial conditions $\Phi_b(\vk)$, where $b=1$ ($b=2$) corresponds to the density (velocity divergence) field, and the line emerging from it carries a wavenumber $\vk$. Lines are time-oriented (with time direction represented by an arrow) and have different indices at both ends, say $a$ and $b$. Each line represents linear evolution described by the propagator $g_{ab}(\etaf-\etai)$ from time $\etai$ to time $\etaf$. Each nonlinear interaction between modes is represented by a vertex, which due to quadratic nonlinearities in the equations of motion is the convergence point of necessarily two incoming lines, with wavenumber say $\vq_{1}$ and $\vq_{2}$, and one outgoing line with wavenumber $\vq=\vq_1+\vq_2$. Each vertex in a diagram then represents the matrix $\gamma_{abc}(\vq,\vq_{1},\vq_{2})$. It is further understood  in Fig.~\ref{DiagramPsiExpansion} that internal indices are summed over and interaction times are integrated over the full interval $[0,\etaf]$.

\begin{figure}
\centerline{\epsfig {figure=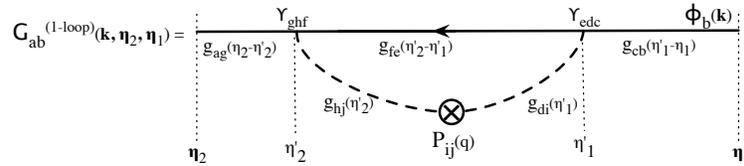,width=10cm}}
\caption{Representation of the one-loop correction to the two-point propagator. The value of this diagram is obtained by the contraction of two incoming lines of $\Psi^{(3)}$ multiplied by the initial power spectrum value. The expression of $G_{ab}^{\oneloop}$ is then obtained after integration over the internal indices $c,\dots, j$, the momentum $q$ and the times $\eta'_{1}$ and $\eta'_{2}$.}
\label{Gab1loop}
\end{figure}

Loop diagrams appear once we calculate statistical averages such as
correlators between fields. An example of such calculation
 (corresponding to the one-loop correction to the linear propagator) is
presented in Fig. \ref{Gab1loop}, where the average over initial conditions is encoded by the dependence on the initial power spectra $P_{ab}^{\init}(q)$, represented by the symbol $\otimes$. 

\begin{figure}
\centerline{\epsfig {figure=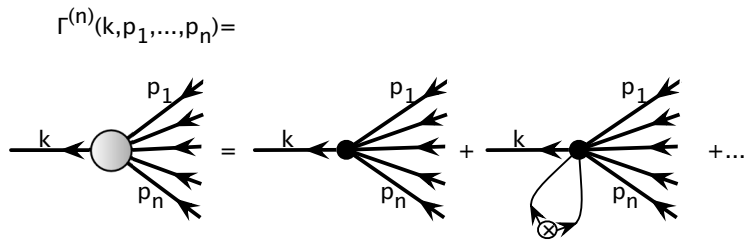,width=10cm}}
\caption{Representation of the first two terms of the multi-point
  propagator $\Gamma^{(n)}$ in a perturbative
  expansion. $\Gamma^{(n)}$ represents the average value of the
  emerging nonlinear mode $\vk$ given $n$ initial modes in the linear
  regime. Here we show the first two contributions: tree-level and
  one-loop. Note that each object represents a collection of
  (topologically) different diagrams  : each black dot represents a set of trees that connect respectively $n+1$ lines for the first term, $n+2$ for the second.}
\label{Gamma5}
\end{figure}

The previous construction can be extended to any number of loops. Note
however that the explicit diagrams to be used depend
  on the statistical properties of the initial fields.  For instance, for Gaussian
  initial conditions the  two-loop diagrams contributing to $G_{ab}(k)$ are obtained from the contraction of 4 incoming lines in the expression of $\Psi^{(5)}$. In case the initial bispectrum is non-vanishing a non-zero contribution can be obtained from the contraction of 3 incoming lines of $\Psi^{(4)}$.

These constructions can be pursued to higher-order propagators. A typical perturbation expansion of $\Gamma^{(n)}$ is presented in Figure~\ref{Gamma5}.

\subsection{The $\Gamma$-expansion}

An important result rigorously shown in \cite{2008PhRvD..78j3521B} and recalled in the introduction, is that the series expansion of the power spectra can be rewritten in term of  product of $\Gamma^{(p)}$ functions as, 
\begin{eqnarray}
\mg\Psi_{a}(\vk_{1})\Psi_{b}(\vk_{2})\md=
\Dirac(\vk_{1}+\vk_{2})
\sum_{p} p! 
\int\dd^3\vq_{1}\dots\dd^3\vq_{p}
\Dirac(\vk_{1}-\vq_{1\dots p})\nonumber\\
\times 
\Gamma^{(p)}_{aa_{1}\dots a_{p}}\left(\vq_{1},\dots,\vq_{p}\right)
\Gamma^{(p)}_{bb_{1}\dots b_{p}}\left(-\vq_{1},\dots,-\vq_{p}\right)\ P^{\init}_{a_{1}b_{1}}(q_{1})\dots P^{\init}_{a_{p}b_{p}}(q_{p}).\label{PkExpansion}
\end{eqnarray}
A further important property of this form is found when primordial metric perturbations are of only adiabatic origin. The initial power spectra then take the form $P_{ab}^{\init}(k)=T_{a}(k)T_{b}(k)P^{\adiab}(k)$
where $P^{\adiab}(k)$ is the primordial power spectrum of the adiabatic modes\footnote{The latter that can be indifferently expressed in terms of the gauge-free potential perturbation or matter density perturbations.}. In this case, using the parity property from Eq.~(\ref{GammapParity}), we  have
\begin{eqnarray}
\mg \Psi_{a}(\vk)\Psi_{b}(\vk')\md
&=&\sum_{p}\frac{1}{p!}
\int\dd^3\vq_{1}\dots \dd^3\vq_{p}
\left[\Gamma^{(p)}_{aa_{1}\dots a_{p}}\left(\vq_{1},\dots,\vq_{p}\right) T_{a_{1}}(q_{1})T_{a_{p}}(q_{p})
\right] 
\nonumber\\
&&\times
\left[\Gamma^{(p)}_{bb_{1}\dots b_{p}}\left(\vq_{1},\dots,\vq_{p}\right) T_{b_{1}}(q_{1})T_{b_{p}}(q_{p})
\right]
P^{\adiab}(q_{1})\dots P^{\adiab}(q_{p}).
\label{PkRecons2}
\end{eqnarray}
In this case, for $a=b$,  the power spectrum  is a sum of squares, that is, of manifestly positive terms. 
As for the RPT case, each of this term will correspond to a ``bump'' contributing to the final power spectra in a limited range of wavemodes. 

The resulting expressions depend on the primordial power spectrum and transfer functions. In this sense the result is a priori model dependent.  
Here, in the context of the $\Gamma$-expansion approach we concentrate on the late time behavior 
of the result thus keeping only the most growing terms. The linear transfer functions are then identical for the 2 components,
$T_{a}(k)=u_{a}T(k)$ with $u_{a}=(1,1)$.
We then simply have
\begin{equation}
P^{\init}_{ab}(k)=u_{a}u_{b}T^2(k)P^{\adiab}(k)=u_{a}u_{b}P_{0}(k)
\end{equation}
and the expression (\ref{PkRecons2}) can be rewritten in,
\begin{equation}
\mg \Psi_{a}(\vk)\Psi_{b}(\vk')\md
=\sum_{p}\frac{1}{p!}
\int\dd^3\vq_{1}\dots \dd^3\vq_{p}\ 
\Gamma^{(p)}_{a}\left(\vq_{1},\dots,\vq_{p}\right)\ \Gamma^{(p)}_{b}\left(\vq_{1},\dots,\vq_{p}\right)
P_{0}(q_{1})\dots P_{0}(q_{p}),
\label{PkRecons3}
\end{equation}
where 
$\Gamma^{(p)}_{a}\left(\vq_{1},\dots,\vq_{p}\right)=\Gamma^{(p)}_{aa_{1}\dots a_{p}}\left(\vq_{1},\dots,\vq_{p}\right)u_{a_{1}}\dots u_{a_{p}}$.
In the following we also assume that the structure of the vertices is that of an Einstein de Sitter universe. As mentioned before, it does not significantly restrict  the validity range of these calculations.

\begin{figure}
\centerline{\epsfig {figure= 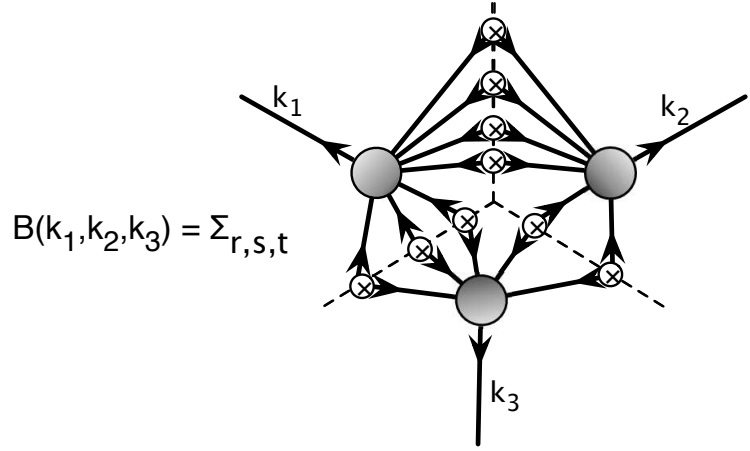,width=10cm}}
\caption{Representation of the resummation rule given by Eq. (\ref{BkExpansion}). For Gaussian initial conditions, the
bispectrum can be seen as a sum of product of $\Gamma^{(p)}$ functions.}
\label{BiSpectreResum}
\end{figure}

This reconstruction scheme for the power spectrum in terms of $\Gamma$-functions can be extended to higher order spectra. For instance, generalizing Eq.~(\ref{PkExpansion}) the bispectrum can be formally rewritten by the resummation,
\begin{eqnarray}
\mg\Psi_{a}(\vk_{1})\Psi_{b}(\vk_{2})\Psi_{c}(\vk_{2})\md&=&
\sum_{r,s,t} \binom{r+s}{r}\binom{s+t}{s}\binom{t+r}{t} r! s! t! 
\int\dd^3\vq_{1}\dots\dd^3\vq_{r}
\ \dd^3\vq'_{1}\dots\dd^3\vq'_{s}
\ \dd^3\vq''_{1}\dots\dd^3\vq''_{t}\ \nonumber\\
&&\,\times
\Dirac(\vk_{1}-\vq_{1\dots r}-\vq'_{1\dots s})\ 
\Dirac(\vk_{2}+\vq'_{1\dots s}-\vq''_{1\dots t})\ 
\Dirac(\vk_{3}+\vq''_{1\dots t}+\vq_{1\dots r})   \nonumber \\
&&\,\times
\Gamma^{(r+s)}_a\left(\vq_{1},\dots,\vq_{r},\vq'_{1},\dots,\vq'_{s}\right) 
\Gamma^{(s+t)}_b\left(-\vq'_{1},\dots,-\vq'_{s},\vq''_{1},\dots,\vq''_{t}\right)\nonumber\\
&&\times\,
\Gamma^{(t+r)}_c\left(-\vq''_{1},\dots,-\vq''_{t},-\vq_{1},\dots,-\vq_{r}\right)
P_{0}(q_{1})\dots P_{0}(q_{r})\ P_{0}(q'_{1})\dots P_{0}(q'_{s})\ P_{0}(q''_{1})\dots
P_{0}(q''_{t}). \nonumber \\
\label{BkExpansion}
\end{eqnarray}
This sum is diagrammatically represented in Fig.~\ref{BiSpectreResum}. We see that it runs over the number of lines that connect each side of the diagram (with the constraint that at most one of the indices $r$, $s$ or $t$ is zero, otherwise we would have a disconnected diagram). The leading order (tree) contribution is then obtained for $r=s=1$, $t=0$ (plus cyclic permutations), up to one-loop corrections (in square brackets) we then have 
\begin{eqnarray}
B(k_1,k_2,k_3)&=& 2\, \Gamma^{(2)}(\vk_1,\vk_2)\, \Gamma^{(1)}(k_1)\, \Gamma^{(1)}(k_2)\, P_0(k_1)P_0(k_2)+ {\rm cyc.} \nonumber \\
 &+&
\Big[ 8 \int d^3q\,  \Gamma^{(2)}(\vk_1-\vq,\vq) \Gamma^{(2)}(\vk_2+\vq,-\vq) \Gamma^{(2)}(\vq-\vk_1,-\vk_2-\vq) P_0(|\vk_1-\vq|) P_0(|\vk_2+\vq|) P_0(q)  \nonumber \\
 &+&
6  \int d^3q\,  \Gamma^{(3)}(-\vk_3,-\vk_2+\vq,-\vq)  \Gamma^{(2)}(\vk_2-\vq,\vq) \Gamma^{(1)}(\vk_3) P_0(|\vk_2-\vq|)P_0(q)P_0(k_3)+ {\rm cyc.}
\Big].
\label{B1loop}
\end{eqnarray}
We will make use of this expression in section \ref{sec:bispectra} below.

\section{Properties of the multi-point propagators}

\subsection{The two-point propagator}

\begin{figure}
\centerline{\epsfig {figure=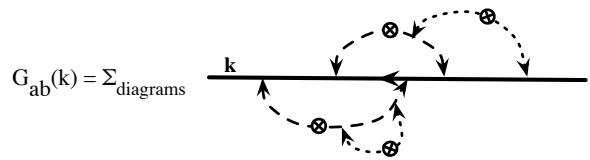,width=8cm}}
\caption{In all the diagrams contributing to $G_{ab}(k)$, as  the one depicted here, there is a line connecting directly the initial time to the final time. This is the principal line, drawn here with a straight thick solid line.   The dominant loops contributing to the resummed propagator are those drawn by dashed lines, while the sub-dominant loops are those in dotted lines.}
\label{PrincipalTrees}
\end{figure}

The large-$k$ behavior of the propagators can be addressed with the help of resummation techniques. This was pioneered in 
\cite{2006PhRvD..73f3520C}, taking advantage that for CDM spectra
there is a characteristic scale set by the rms displacement (or
velocity) field that sets the typical momenta inside loop diagrams,
thus by large-$k$ we mean specifically $k$ larger than this
characteristic scale. This idea was put in a more  general footing in \cite{2011arXiv1109.3400B} where the concept of the eikonal 
approximation is introduced.  In this context it is possible to compute the expression of the non-linear propagators taking into account 
the full resummed contributions of modes $q\ll k$. 

The resulting expression of the propagator is then valid in the high $k$ limit.
More specifically one finds that
\begin{equation}
G_{ab}(k,\etaf,\etai)\ \ \to\ \ g_{ab}(\etaf-\etai)\exp\left(-\frac{k^2\sigmadv(\etaf,\etai)}{2}\right)\ 
\label{Gabloops}
\end{equation}
where $\sigmav(\etaf,\etai)$ is the r.ms. of the one-point displacement field, $\vd(\etaf,\etai)$, from time $\etai$ to time $\etaf$. More precisely
the latter is given by
\begin{equation}
\vd(\etaf,\etai)=\int_{\etai}^{\etaf}\dd \eta \int\dd^3\vq\frac{\vq}{q^2}\frac{\theta(\vq,\eta)}{f_{+}(\eta)\mH}
\end{equation}
assuming the velocity field is potential. The functional form (\ref{Gabloops}) is valid assuming the large scale displacement field obeys a Gaussian statistics.  In that case the exponential damping is entirely determined by the variance of the displacement along one direction,
\begin{equation}
\sigmadv(\etaf,\etai)=\frac{1}{3}\langle\vd^2(\etaf,\etai) \rangle.
\end{equation} 
In case the displacement is given by its linear expression and assuming it is dominated by the growing mode contribution
one then has, 
\begin{equation}
\sigmadv(\etaf,\etai)=(e^\etaf-e^\etai)^2\,\int\frac{\dd^{3} \vq}{3q^{2}}\ P_{0}(q),\label{sigmavdef}
\end{equation} 
where $P_{0}(q)$ is the initial linear power spectrum. 
This result was originally derived in \cite{2006PhRvD..73f3520C} from the explicit contribution of a large subset of diagrams - those 
that are directly connected to the principal line  (see
  Fig.~\ref{PrincipalTrees}). The eikonal approximation shows that this result is very general. It is valid in particular irrespectively of the time dependence of the velocity field.
As shown in \cite{2011arXiv1109.3400B},
this construction amounts to compute the displacement field from its
linear expression.  It is possible to include corrections to the
  displacement field statistical properties beyond linear theory. This
  was noticed in \cite{2011JCAP...06..015A} where 1-loop corrections
  to the variance of the  displacement field are included in the
  calculation of the propagator damping function. It should be  noted however that whenever the displacement field is not Gaussian distributed, the damping factor is not a function of its variance only. 
This can be naturally be taken into account in the eikonal
approximation. For instance, the standard results can be extended to
models with primordial non-Gaussian initial conditions for which one
can  recover both the resummation results and the $\Gamma$-expansion
formulae (see \cite{2010PhRvD..82h3507B} for details). In this case the exponential factor is replaced by
\begin{equation}
\exp\left(-\frac{k^2\sigmadv(\etaf,\etai)}{2}\right)\ 
\ \ \to\ \ \
\exp\left(-\sum_{p=2}^{\infty}\frac{\langle(\vd(\etaf,\etai)\cdot\vk)^p\rangle_{c}}{p!}\right).
\label{fkexpression}
\end{equation}
In this expression the variance  of the 1D displacement field along
$\vk$  has been replaced by the whole cumulant series of the 1D displacement field. 
{Note this form can be extracted from another route, based on the use of Lagrangian space variables \cite{2008PhRvD..77f3530M,2011PhRvD..83h3518M}. In this case, however, that it corresponds to the leading behavior in the high-$k$ limit is ambiguous.}

On the other hand, as stressed in the introduction, the nonlinear expression of $G_{ab}$ can be approached with a perturbative series expansion. Formally one can write 
$G_{ab}$ as,
\begin{equation}
G_{ab}(k,\etaf,\etai)=g_{ab}(\etaf-\etai)+\delta G_{ab}^{\oneloop}(k,\etaf,\etai)+\delta G_{ab}^{\twoloops}(k,\etaf,\etai)+\dots
\label{GabSeries}
\end{equation}
where successive loop corrections are included. It is known  that $G_{ab}^{\oneloop}(k,\etaf,\etai)$ behaves like 
$-{k^2\sigmadv(\etaf,\etai)}/2$, etc.,  when $k$ is large so that the perturbative expansion of (\ref{Gabloops})
and (\ref{GabSeries}) agree when $k$ is large. And this is expected to be true at all order in perturbation theory. 
This is actually the meaning one can give to the limit written in (\ref{Gabloops}). Note that sub-leading terms of (\ref{Gabloops}) are obviously expected to be present in the expression of $G_{ab}(k,\etaf,\etai)$. If they appear within the exponential they would change  the 
normalization factor.

In all cases what we expect is that (\ref{Gabloops}) captures the large $k$ behavior of the propagator whereas the expansion
(\ref{GabSeries}) is expected to be a precise description of its low $k$ behavior. One of the objectives of this paper is explore how these two limit behaviors can be matched together. 

\subsection{The RPT  interpolation scheme for the two-point propagator}

The one-loop expressions of the two-point propagator have been
explicitly calculated in \cite{2006PhRvD..73f3520C}. We summarize
them in this section,  and their interpolation to the resummed high-$k$ limit,  as it will be useful to compare to our new proposal in section IV.

The computation of the one-loop contribution involves in general different time-dependent functions. They are all of the form $a^{\nu}$ where $\nu$ is an integer or a half-integer. This is a consequence of the structure of the time dependence of the linear propagator and of the fact that we
assume $\Omega_{ab}$ to be time independent. For each time dependence, each component $G_{ab}^{\oneloop}$ has 
a specific $k$ dependence that can be computed. But although there are 6 different $a^{\nu}$ functions that intervene
in the expression of the one-loop diagram, the whole result can be collected into only 4 different $k$-dependent functions~\cite{2006PhRvD..73f3520C}. 
We recall here the explicit expressions
of those results. One obtains,
\begin{eqnarray}
\delta G_{11}^{\oneloop}(k,a)&=& \frac{3}{5} \alpha(a)\ f(k) - \frac{3}{5} \beta(a)\ i(k) - \frac{2}{5} \gamma(a)\ h(k) + \frac{2}{5} \delta(a)\ g(k), \nonumber \\
\delta G_{12}^{\oneloop}(k,a)&=& \frac{2}{5} \alpha(a)\ f(k) - \frac{2}{5} \beta(a)\ h(k) + \frac{2}{5} \gamma(a)\ h(k) - \frac{2}{5} \delta(a)\ f(k), \nonumber \\
\delta G_{21}^{\oneloop}(k,a)&=& \frac{3}{5} \alpha(a)\ g(k) - \frac{3}{5} \beta(a)\ i(k) + \frac{3}{5} \gamma(a)\ i(k) - \frac{3}{5} \delta(a)\ g(k), \nonumber \\
\delta G_{22}^{\oneloop}(k,a)&=& \frac{2}{5} \alpha(a)\ g(k) - \frac{2}{5} \beta(a)\ h(k) - \frac{3}{5} \gamma(a)\ i(k) + \frac{3}{5} \delta(a)\ f(k), 
\label{proponeloop1}
\eeqa
with 
\beqa
\alpha(a)=a^3-\frac{7}{5}a^{2}+\frac{2}{5}a^{-1/2},\,\beta(a)=\frac{3}{5}a^2-a+\frac{2}{5}a^{-1/2},\,\gamma(a)=\frac{2}{5}a^{2}-a^{1/2}+\frac{3}{5}a^{-1/2},\,\delta(a)=\frac{2}{5}a^{2}-\frac{7}{5}a^{-1/2}+a^{-3/2}.
\label{functionstime}
\nonumber
\end{eqnarray}
and
\begin{eqnarray}
f(k)&=&\int\frac{1}{504 k^3 q^5}
\left[6k^7q-79k^5q^3+50q^5k^3-21kq^7+\frac{3}{4}(k^2-q^2)^3(2k^2+7q^2)\ln \frac{|k-q|^2}{|k+q|^2}\right]P_0(q) \, \dd^3\vq, \nonumber \\
g(k)&=&\int\frac{1}{168k^3q^5}
\left[6k^7q-41k^5q^3+2k^3q^5-3kq^7+\frac{3}{4}(k^2-q^2)^3(2k^2+q^2)\ln\frac{|k-q|^2}{|k+q|^2}\right]P_0(q)\, \dd^3\vq,     \nonumber \\
h(k)&=&\int\frac{1}{24k^3q^5}
\left[6k^7q+k^5q^3+9kq^7+\frac{3}{4}(k^2-q^2)^2(2k^4+5k^2q^2+3q^4)\ln\frac{|k-q|^2}{|k+q|^2}\right]P_0(q) \,\dd^3\vq,   \nonumber    \\
i(k)&=&\int\frac{-1}{72k^3\,q^5}
\left[6k^7q+29k^5q^3-18k^3q^5+27kq^7+\frac{3}{4}(k^2-q^2)^2(2k^4+9k^2q^2+9q^4)\ln\frac{|k-q|^2}{|k+q|^2}\right] P_0(q) \, \dd^3\vq,  \nonumber \\    
\label{functionsofk}
\end{eqnarray}
We can notice that all these functions satisfy~\cite{2006PhRvD..73f3520C}
\begin{equation}
f(k),\ g(k),\ h(k),\ i(k)\to -\frac{k^2}{2}\int\frac{\dd^3\vq}{3q^2}P_{0}(q)
\label{fghilimit}
\end{equation}
when $k$ is large. The time dependent functions also obey remarkable properties,
\begin{eqnarray}
\alpha(a)-\beta(a)&=&a(a-1)^{2},\label{alphambeta}\\
\delta(a)-\gamma(a)&=&a^{-3/2}(a-1)^{2},\label{deltamgamma}
\end{eqnarray}
so that, in the high $k$ limit we indeed have
\begin{equation}
\delta G_{ab}^{\oneloop}(k,\etaf,\etai)\ \to -\frac{1}{2}k^2\sigmadv(\etaf,\etai)\  g_{ab}(\etaf,\etai).
\end{equation}

One can also remark that $\alpha(a)$ is the most rapidly growing function and is therefore expected to dominate at late times. In this
case only the functions $f(k)$ and $g(k)$ play a role in the expression of the propagators.
Irrespectively of this limit, the proposed exponentiation scheme in \cite{2006PhRvD..73f3520C} is the following. It is based on the exponentiation of terms either identified as the growing modes or the decaying modes,
\begin{eqnarray}
g_{11}+\delta G^{(1)}_{11}\,\rightarrow\,G_{11}(k,a)&=&\frac{3}{5} \, a \, {\rm e}^{(\alpha_{g}(a) f(k) - \beta_g(a) i(k))} + \frac{2}{5} \, a^{-3/2} \, {\rm e}^{(\delta_{d}(a) g(k) - \gamma_d(a) h(k))} \nonumber ,\\
g_{12}+\delta G^{(1)}_{12}\,\rightarrow\,G_{12}(k,a)&=&\frac{2}{5} \, a \, {\rm e}^{(\alpha_{g}(a) f(k) - \beta_g(a) h(k))} - \frac{2}{5} \, a^{-3/2} \, {\rm e}^{(\delta_{d}(a) f(k) - \gamma_d(a) h(k))}  \nonumber ,\\
g_{21}+\delta G^{(1)}_{21}\,\rightarrow\,G_{21}(k,a)&=&\frac{3}{5} \, a \, {\rm e}^{(\alpha_{g}(a) g(k) + \gamma_g(a) h(k))} - \frac{3}{5} \, a^{-3/2} \, {\rm e}^{(\delta_{d}(a) g(k) + \beta_d(a) i(k))} \nonumber ,\\
g_{22}+\delta G^{(1)}_{22}\,\rightarrow\,G_{22}(k,a)&=&\frac{2}{5} \, a \, {\rm e}^{(\alpha_{g}(a) g(k) - (3/2) \gamma_g(a) i(k))} + \frac{3}{5} \, a^{-3/2} \, {\rm e}^{(\delta_{d}(a) f(k)-(2/3) \beta_d(a) h(k))}.
\label{model3one}
\end{eqnarray}
where we have redefined the $\alpha$ -- $\gamma$ functions in Eq.~(\ref{functionstime}) through  
$a\,\alpha_{g}(a)=\alpha(a)$, $a\,\beta_g(a)=a^{-3/2} \beta_d(a)=\beta(a)$, $a\,\gamma_g(a)=a^{-3/2} \gamma_d(a)=\gamma(a)$
and $a^{-3/2}\,\delta_{d}(a)=\delta(a)$. At one-loop order these forms agree with the results (\ref{proponeloop1}). They also agree with the limit form (\ref{Gabloops}) because of the properties (\ref{fghilimit}-\ref{deltamgamma}). 
These forms also present a number of key properties: they are decaying functions of time and of $k$. This is ensured in particular by the fact that the terms under the exponential is always negative. Note that there are no free parameters in this construction: given an initial spectrum and cosmological parameters those form fully predict $G_{ab}(k,\eta)$. 
These forms are the nonlinear propagator used in the RPT formalism. They have been successfully tested against  N-body simulations so alternative interpolation schemes cannot significantly depart from them. 

It  should be remarked however that those constructions do not necessarily represent the unique possible interpolation scheme. In particular if one allows the possibility of adding more than two exponentials, then one would obtain a whole set of alternative formulations. Before we move on to the formulation we propose in this paper let us first describe the multi-point propagator results.

\subsection{The multi-point propagators}

\begin{figure}[t!]
\centerline{\epsfig {figure=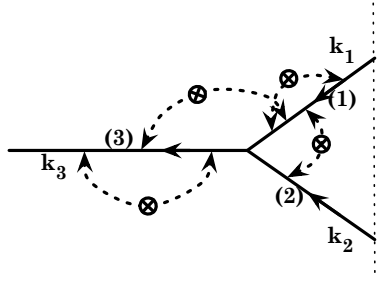,width=5cm}}
\caption{Example of dominant loop contributions for the three-point propagator  $\Gamma^{(2)}_{abc}(\vk_{1},\vk_{2},\vk_{3})$. The final expression is obtained by the sum of an infinite number of such loops and over all possible interaction times.} 
\label{Gamma2Loops}
\end{figure}

Let us continue with the diagrammatic approach, extended to multi-point propagators.  
The concept of principal line can be extended to the multi-point propagators. One can then define a ``principal tree'' which corresponds to the diagram when the propagator is taken at tree order (starting with order $4$ there might be more than one possible tree). The diagrams contributing to the high-$k$ limit of the propagator are those that are directly connected to the principal tree~\cite{2008PhRvD..78j3521B}.

For a given tree shape $(q)$ (for instance, one of the diagrams of Fig.~\ref{Gamma2Loops}), a careful resummation of all these diagrams 
gives the following result (for Gaussian initial conditions),
\begin{equation}
\Gamma^{(p-q)}_{a b_{1}\dots b_{p}}(\vk_{1},\dots,\vk_{p},\etaf,\etai,\eta'_{1},\dots,\eta'_{p-1})\ \ \to\ \ \exp\left(-\frac{k_{1\dots p}^2\sigmadv(\etaf,\etai)}{2}\right)\ 
\Gamma^{(p-q),{\rm tree}}_{a b_{1}\dots b_{p}}(\vk_{1},\dots,\vk_{p},\etaf,\etai,\eta'_{1},\dots,\eta'_{p-1}).
\label{Gammapqloops}
\end{equation}
where $(q)$ labels a given topology and $\eta'_{i}$ are the time values at each vertex position.
As this result is valid for any topology and  any time, after proper summation we simply have, 
\begin{equation}
\Gamma^{(p)}_{a b_{1}\dots b_{p}}(\vk_{1},\dots,\vk_{p},\etaf,\etai)\ \ \to\ \ \exp\left(-\frac{k_{1\dots p}^2\sigmadv(\etaf,\etai)}{2}\right)\ 
\Gamma^{(p),{\rm tree}}_{a b_{1}\dots b_{p}}(\vk_{1},\dots,\vk_{p},\etaf,\etai).
\label{Gammaploops}
\end{equation}
This result generalizes the one for the  two-point propagator. Note that this result can also be derived in the context of the eikonal
approximation showing that $\sigmav(\etaf,\etai)$ does not need to be computed in the linear regime.

 Similarly to the two-point propagator, it is possible to
obtain the low-$k$ behavior of the multi-point propagators by
perturbative series expansion,
\begin{eqnarray}
\Gamma^{(p)}_{a b_{1}\dots b_{p}}(\vk_{1},\dots,\vk_{p},\etaf,\etai)&=&
\Gamma^{(p),{\rm tree}}_{a b_{1}\dots b_{p}}(\vk_{1},\dots,\vk_{p},\etaf,\etai)+
\Gamma^{(p),\oneloop}_{a b_{1}\dots b_{p}}(\vk_{1},\dots,\vk_{p},\etaf,\etai)\nonumber\\
&&+\Gamma^{(p),\twoloops}_{a b_{1}\dots b_{p}}(\vk_{1},\dots,\vk_{p},\etaf,\etai)+\dots
\end{eqnarray}
Note that in this case, even for  the late-time behavior of the one-loop corrections, the relative sign between the tree term and the one-loop term is not fixed. It depends on both the geometry and the amplitude of the modes.

Again, the aim of this paper is to propose an interpolation scheme between the known large and small scale asymptotic that fully respects these two limits.

\section{Proposed interpolation scheme}

\subsection{The case of the most growing mode}

The construction of our matching scheme is  based on the analysis of the structure of the multi-loop terms
corrections.  To start with let us concentrate our presentation on the late time behavior of the propagator~\footnote{In the context of the $\Gamma$-expansion approach it is in practice needless to consider sub-dominant terms and furthermore the extension of this construction to the general time
does not introduce any new difficulty.}. In this limit it is legitimate to assume that the initial fields are in growing mode only. We are then left with two independent quantities for the propagator,
\begin{equation}
G_{a+}=G_{a1}+G_{a2}.
\end{equation}
Up to one-loop order the late time expression of $G_{a+}$ is,
\begin{equation}
G_{a+} = e^{\eta}+e^{3\eta} f_{a}(k)
\end{equation}
where $f_{a}(k)$ is either $f(k)$, for $a=1$, or $g(k)$ for $a=2$. The large $k$ limit 
of $f_{a}(k)$ is $f_{a}(k) \to -k^2 \sigmadv/2$. The RPT expression of the propagator is then 
\begin{equation}
G_{a+}(k,\eta) = e^{\eta} \exp[e^{2\eta} f_{a}(k)].\label{CS1}
\end{equation}

\begin{figure}[t]
\begin{center}
\epsfig{file=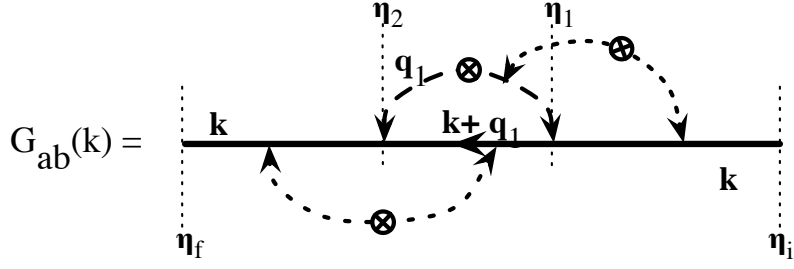,scale=0.8}
\caption{Large $\vk$, $\vq_{1}$ regularization of the one-loop diagram of the two-point propagator by higher order loops corrections in the high-$k$ limit. The main loop (in dashed line) is to be computed when $\vq$ is in the UV domain. }
\label{OneLoopReg}
\end{center}
\end{figure}

\begin{figure}[htbp]
\begin{center}
\epsfig{file=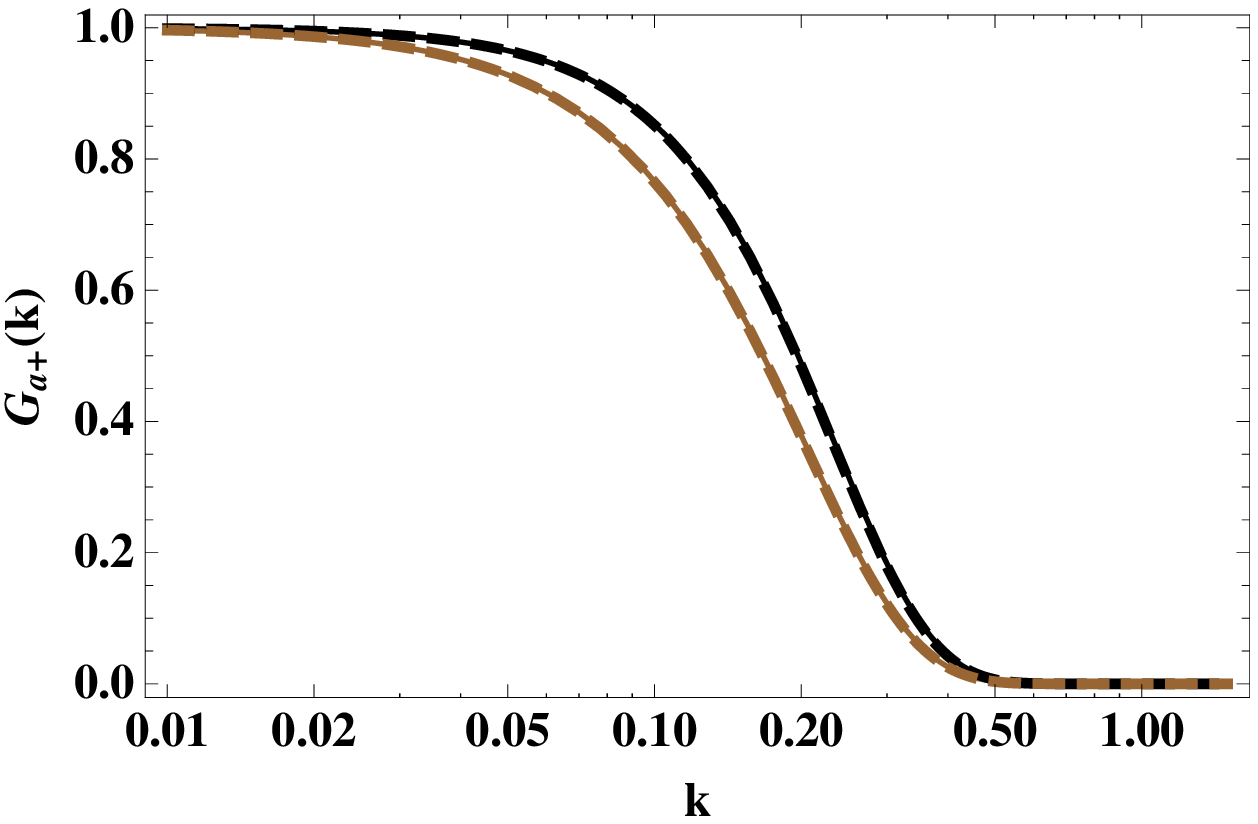,height=5cm}\hspace{1cm}\epsfig{file=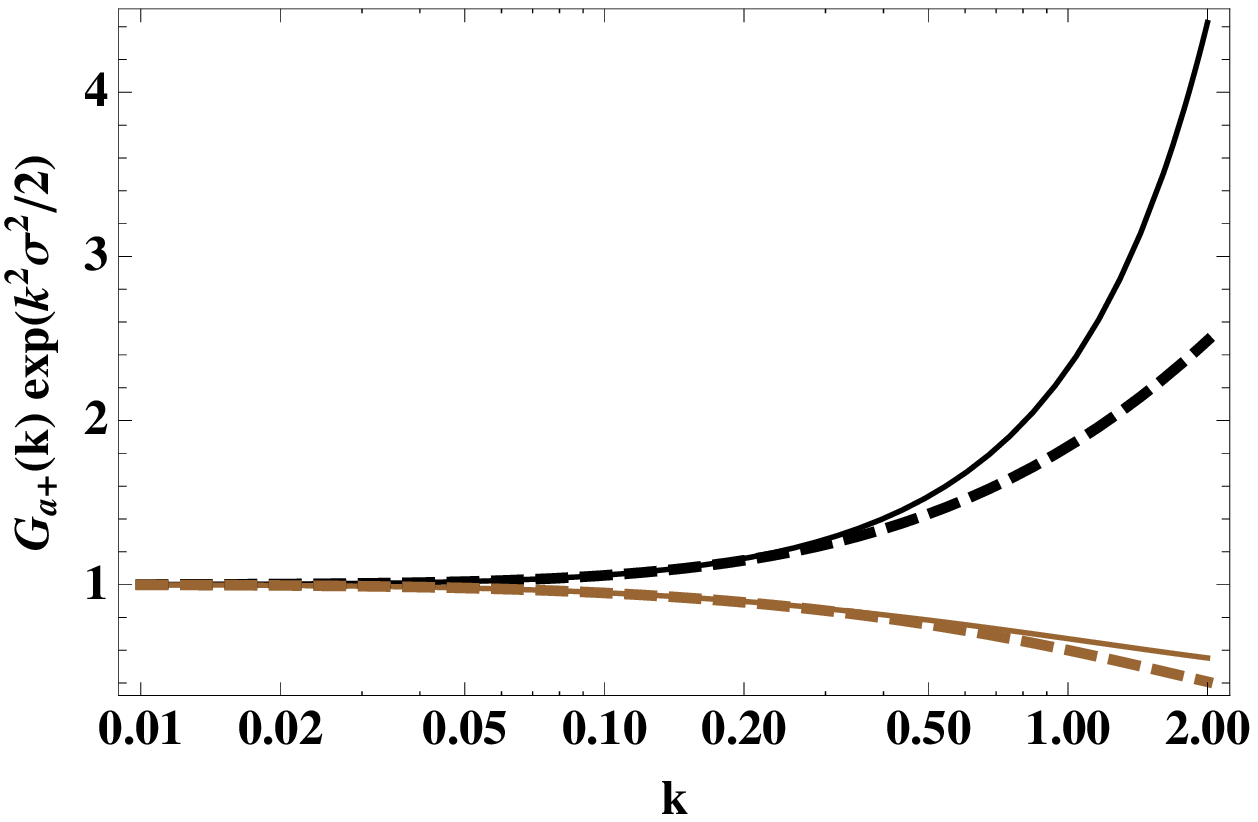,height=5cm}
\caption{Comparison of prescriptions for exponentiation schemes for the density (black upper lines) and velocity (brown lower lines) fields. The thin solid lines correspond to the RPT prescription, the thick dashed lines to this work. The left  panel shows the propagators, the right corresponds to the propagator when they are divided by the global damping factor $\exp(-k^2\sigmadv/2)$. The computations are made for $z=0$ and for a $\Lambda$-CDM cosmology (see Section \ref{sec:simulations} for details).}
\label{Gaplus}
\end{center}
\end{figure}

The alternative prescription we propose is based on the following observation regarding the renormalization of the one-loop result: let us consider the diagram on Fig.~\ref{OneLoopReg} where the intermediate times $\eta'_{1}$ and $\eta'_{2}$ are fixed and  the value of $\vq_{1}$ is also fixed such that its norm is large. The crucial observation we now make is that this object is then nothing but $\Gamma^{(3)}(\vq_{1},-\vq_{1},\vk,\etaf,\etai)$, where the three incoming modes from the initial conditions correspond  to the two dashed lines (joined at the initial power spectrum, i.e. the $\otimes$ symbol) and the rightmost $k$-mode. But we  now have a good understanding of its renormalization properties, given by Eq. (\ref{Gammapqloops}). This equation tells us that  we know how to resum all its loop corrections (some of which are represented by dotted loops in Fig.~\ref{OneLoopReg}) in the large $q_{1}$ and $k$ limit. Because it corresponds to the effects of long wavemodes, let us call this the infra-red (IR) correction. The expression we find is an application of the general result for multi-point propagator and it leads to a simple $\exp(-k^2\sigmadv/2)$ factor. It is important to note that it is independent of $q_{1}$ (and intermediate times $\eta_{1}$ and $\eta_{2}$). We still have to perform the integral $\eta_{1}$ and $\eta_{2}$ and then over $q_{1}$. For the latter however we have to bear in mind that it cannot run over all possible values: it has to avoid its IR part. We can then split this integral into 2 parts, one IR, for which this result is not valid, and one UV for which it applies. Then the value of that set of diagrams would  be
\begin{equation}
\delta G_{a+}=e^{3\eta} f_{a}^{(IR)}(k)+e^{3\eta} f_{a}^{(UV)}(k)\exp\left(-k^2\sigmadv/2\right).
\end{equation}
We then note that the first term is simply the first (non trivial) term of the usual IR resummation of diagrams, $\exp(-k^2\sigmadv/2)$. 
We are then let to simply set
\begin{eqnarray}
f_{a}^{(IR)}(k)&=&-\frac{1}{2}k^2\sigmadv.\\
f_{a}^{(UV)}(k)&=&f_{a}(k)-f_{a}^{(IR)}(k).
\end{eqnarray}
This identification leads to the following form,
\begin{equation}
G_{a+}^{\reg}  = e^{\eta}+\delta G_{a+}=e^{\eta} \left(1+e^{2\eta} f_{a}(k)+  k^2 \sigmadv/2\right) \exp\left(- k^2 \sigmadv/2\right)
\end{equation}
for a ``regularized'' propagator
which compared to the expression (\ref{CS1}) amounts to replacing 
\beq
\exp\left(e^{2\eta}f_{a}(k)+k^2 \sigmadv/2\right) \rightarrow (1+e^{2\eta} f_{a}(k)+ k^2 \sigmadv/2). 
\eeq
Note that these quantities are both finite at large $k$. In Fig. \ref{Gaplus}, we compare these two prescriptions for the
density and velocity fields at $z=0$ for a $\Lambda$-CDM cosmology. They are virtually
indistinguishable when one considers the propagator shape. They
significantly depart from one another only for $k\simeq 1 \Mpc$, as shown 
on the right panel, that is at scales where the exponential damping is already extremely strong.

\subsection{The general prescription}

Since the resummation properties of the $\Gamma$ functions preserve the topology and the intermediate time structure, the whole procedure applies to the full time dependence of the one-loop term. The integral splitting can then be done more generally and one gets,
\begin{eqnarray}
G_{ab}^{\reg}(k,\etaf,\etai)&=&\left[g_{ab}(\etaf,\etai)+\delta G_{ab}^{\oneloop}(k,\etaf,\etai)+\frac{1}{2}k^{2}\sigmadv(\etaf,\etai) g_{ab}\right]\exp\left(-\frac{k^2\sigmadv(\etaf,\etai)}{2}\right).
\end{eqnarray}
The relation with the RPT proposed forms given by Eqs. (\ref{model3one}) is here more complicated. It is clear however that both propositions
agree at the one-loop correction level and both propositions exhibit the same high-$k$ behavior.

One important aspect of this construction, which will be exploited in the following, is that it can obviously be extended to multipoint propagators,
\begin{eqnarray}
\Gamma^{\reg}_{ab_{1}\dots b_{p}}(\vk_{1},\dots,\vk_{p},\etaf,\etai)&=&
\left[\Gamma^{{\rm tree}}_{ab_{1}\dots b_{p}}(\vk_{1},\dots,\vk_{p},\etaf,\etai)+\delta \Gamma^{\oneloop}_{ab_{1}\dots b_{p}}(\vk_{1},\dots,\vk_{p},\etaf,\etai)+\right.\nonumber\\
&&\hspace{-2cm}\left.+\frac{1}{2}k^{2}\sigmadv(\etaf,\etai)  \Gamma^{{\rm
      tree}}_{ab_{1}\dots
    b_{p}}(\vk_{1},\dots,\vk_{p},\etaf,\etai)\right]\exp\left(-\frac{k^2\sigmadv(\etaf,\etai) }{2}\right)
\label{regGammaponeloop}
\end{eqnarray}
where $k=\vert\vk_{1}+\dots+\vk_{p}\vert$. 
By construction this form is such that it has both the correct one-loop correction and the correct large-$k$ behavior. One can also remark that, unlike the RPT interpolation proposal where the matching is done in the ``basis" given by density and velocity fields, the resulting form here is  independent on the basis chosen to represent the cosmic fluids. This may not be an important difference  for the CDM-only case  but could be of importance when one wishes to extend the field content to other degrees of freedom (e.g. the case where one has in addition 
baryons, massive neutrinos, extra fields in the context of modified gravity theories, etc.) 

Finally, another important advantage of this construction is that it
can be pursued to any order in loop corrections  in a rather
straightforward way,
\begin{equation}
\Gamma^{\reg}_{ab_{1}\dots b_{p}}=
\left[\Gamma^{{\rm tree}}_{ab_{1}\dots b_{p}}+\delta \Gamma^{\oneloop}_{ab_{1}\dots b_{p}}+\frac{1}{2}k^{2}\sigmadv \Gamma^{{\rm tree}}_{ab_{1}\dots b_{p}}+\delta \Gamma^{\twoloops}_{ab_{1}\dots b_{p}}+\ct\right]\exp\left(-\frac{k^2\sigmadv}{2}\right)
\label{regGammap}
\end{equation}
where $\ct$ is a counter-term such that the 2-loop expression of the reg. expression is exact,
\begin{equation}
\ct= - \frac{1}{2}\left(\frac{k^2\sigmadv}{2}\right)^2\Gamma^{{\rm tree}}_{ab_{1}\dots b_{p}}
+\frac{k^2\sigmadv}{2}\ \delta \Gamma^{\oneloop}_{ab_{1}\dots b_{p}}
\end{equation}

\begin{figure}[ht]
\begin{center}
\epsfig{file=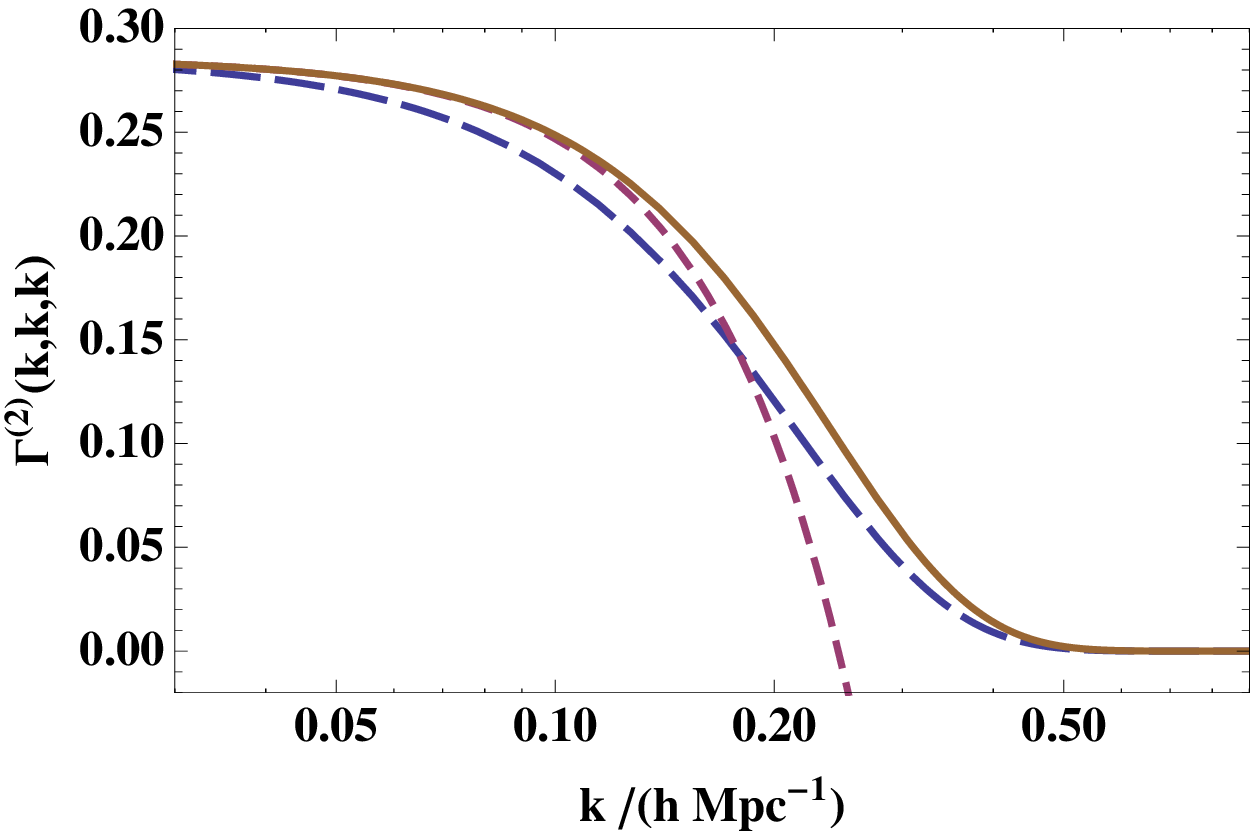,height=5cm}\hspace{1cm}\epsfig{file=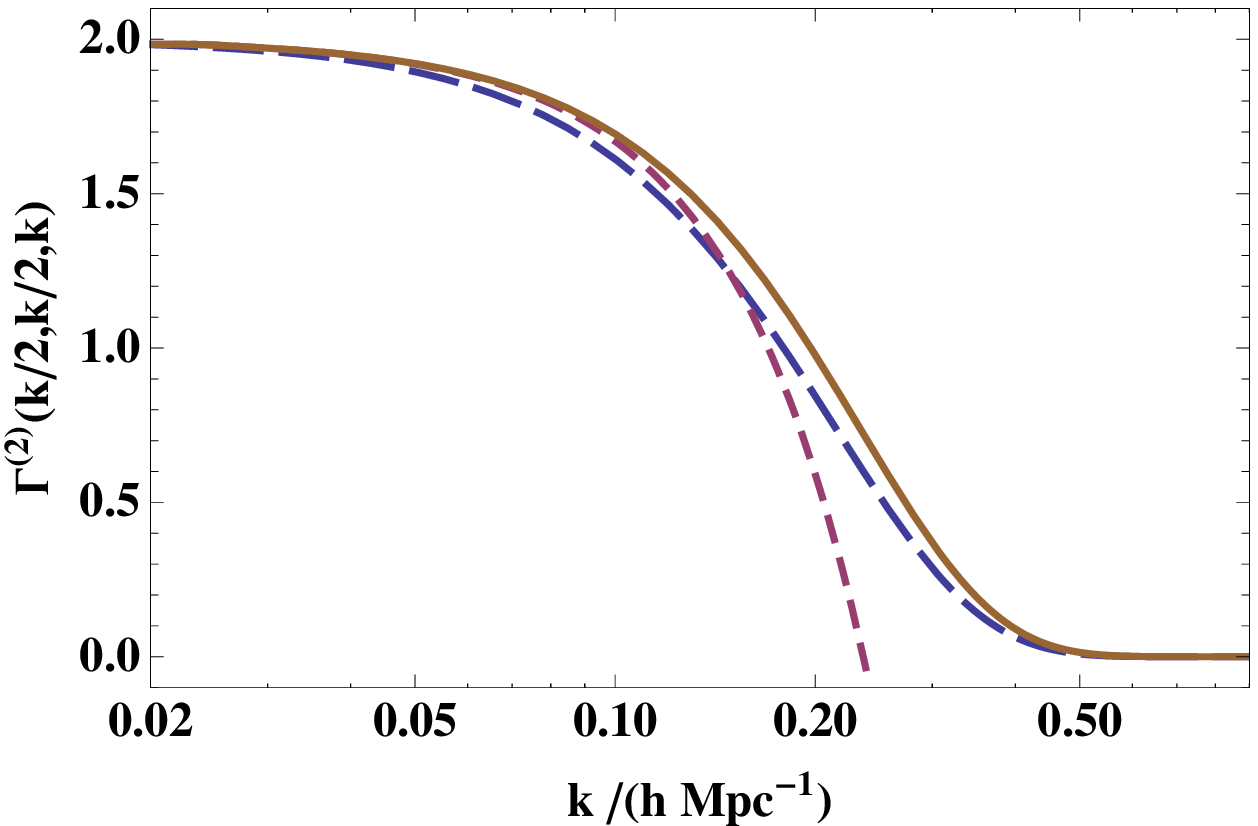,height=5cm}\\
\epsfig{file=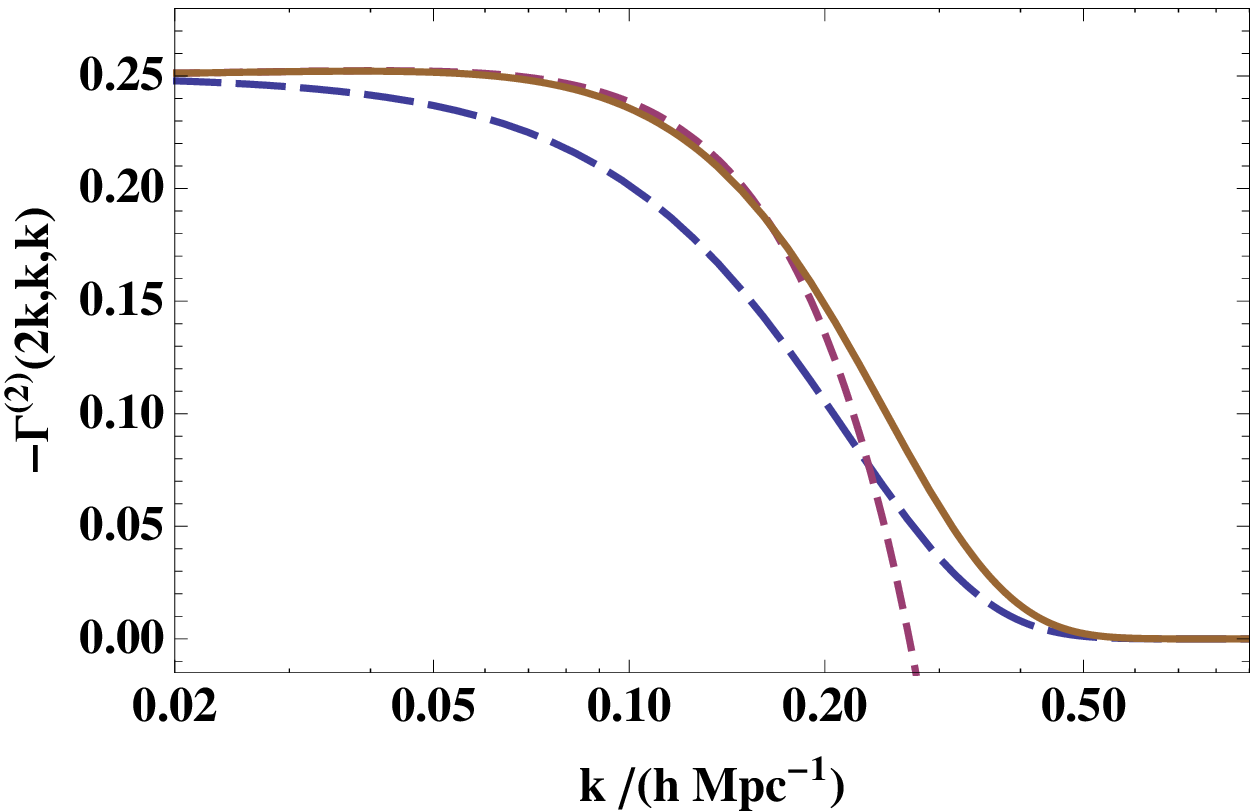,height=5cm}\hspace{1cm}\epsfig{file=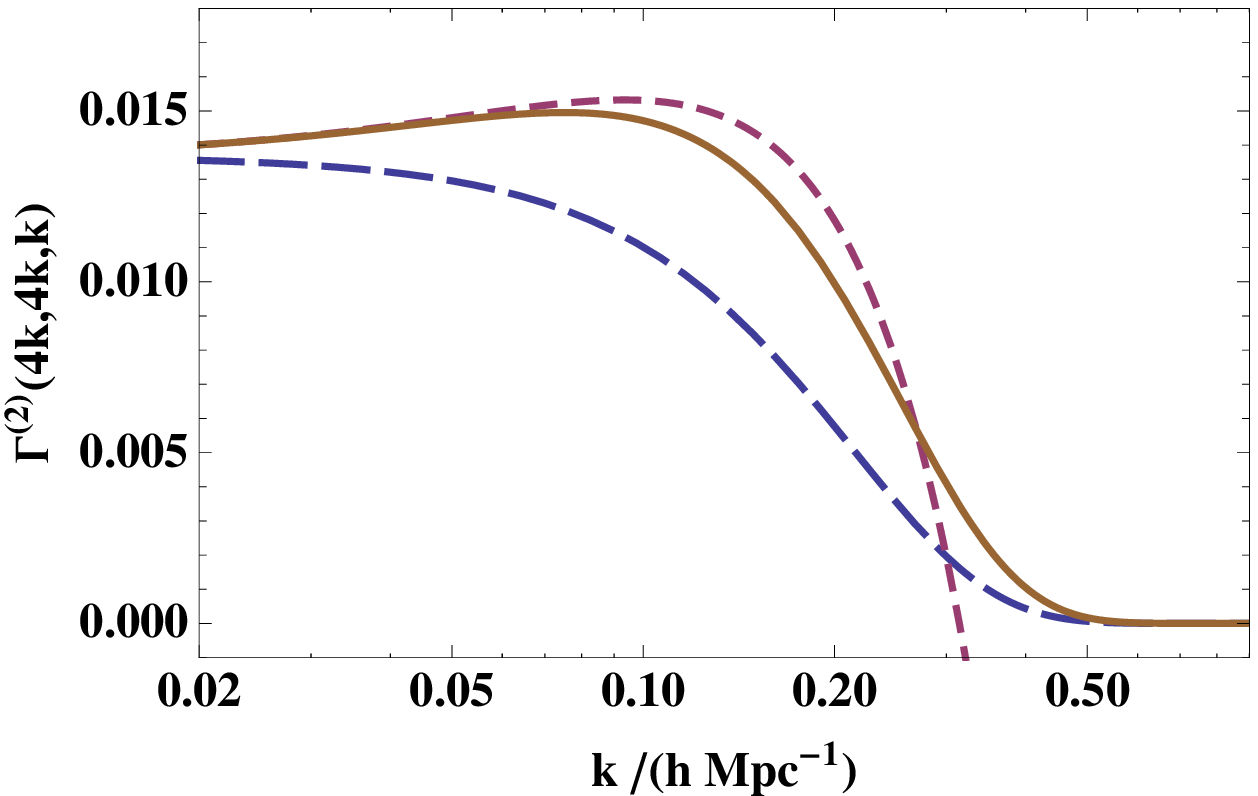,height=5cm}
\caption{The shape of the $\Gamma^{(2)}$ propagator for ``equilateral''
  ($k_{1} = k_{2} = k_{3}=k$, top left), ``colinear'' ($k_{1} = k_{2} =
  k/2$ ; $k_{3} = k$, top right), ``anti-colinear'' ($k_{1} =  2k$
  ; $k_{2}=k_{3} = k$, bottom left) and ``squeezed'' ($k_{1} = k_{2}=4k$ ;
  $k_{3} = k$, bottom right).  The  long dashed (blue) lines
  correspond to the simple exponential cut-off obtained in the {\it high-k}
  limit, Eq.~(\ref{Gammaploops}), the short dashed (violet) lines are the tree plus one-loop order
  results and the solid (brown) line correspond to the interpolation scheme proposed this paper. These comparisons are made for a power spectrum corresponding to $\Lambda-$CDM model at $z=0$.}
\label{G2reg}
\end{center}
\end{figure}

Let us now illustrate this with some examples. The resulting interpolation scheme for $\Gamma^{(2)}$ for specific geometries is shown in Fig. \ref{G2reg}
for its late time behavior, showing that the interpolation is rather smooth. In particular it can handle the fact that tree-order and one-loop corrections have different signs. This is the case illustrated  in the lower right panel in Fig. \ref{G2reg}.

\subsection{The case of Non-Gaussian initial conditions}

\begin{figure}[ht]
\begin{center}
\epsfig{file= 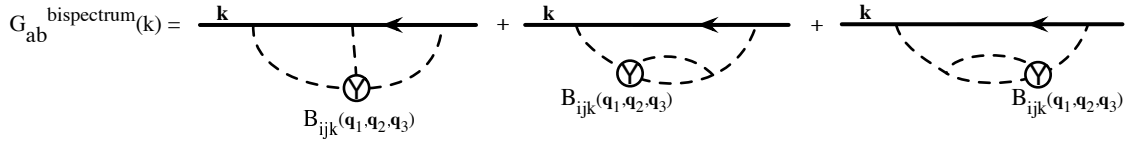,width=15cm}
\caption{Dominant complementary diagrams contributing to the one-point propagator in case of PNGs. They make intervene  
the primordial bispectrum $B_{abc}(q_{1},q_{2},q_{3})$. For models of cosmological interest they are of the order of the 2-loop 
contributions. When those diagrams are computed in the eikonal approximation, that is when the vertices are approximated by their
high $k$ limit they all vanish for parity reasons. They do not so in general.}
\label{NGInitConds}
\end{center}
\end{figure}

The case of PNGs can similarly be taken into account. In this case the damping factor is changed in order to take into account 
the higher order cumulants of the 1D displacement field as given in Eq. (\ref{fkexpression}). 
Still it is possible to apply the same regularization scheme except that the counter terms have to be recomputed.

Novel two-loop order terms, depicted on Fig. \ref{NGInitConds},  are due to the primordial bispectrum. In the eikonal approximation (e.g. when the vertex values are computed for $q_{i}\ll k$), these diagrams vanish however. They therefore do not have counter terms in the regularization scheme we propose.  Differences in the counter terms appear only at the three-loop order. It corresponds to the fact that
the damping factor given by Eq. (\ref{fkexpression}) is not sensitive to the primordial bispectrum, but it is to the primordial trispectrum.

\section{Comparison with numerical simulations}
\label{sec:simulations}

\begin{figure}[ht]
\begin{center}
\epsfig{file=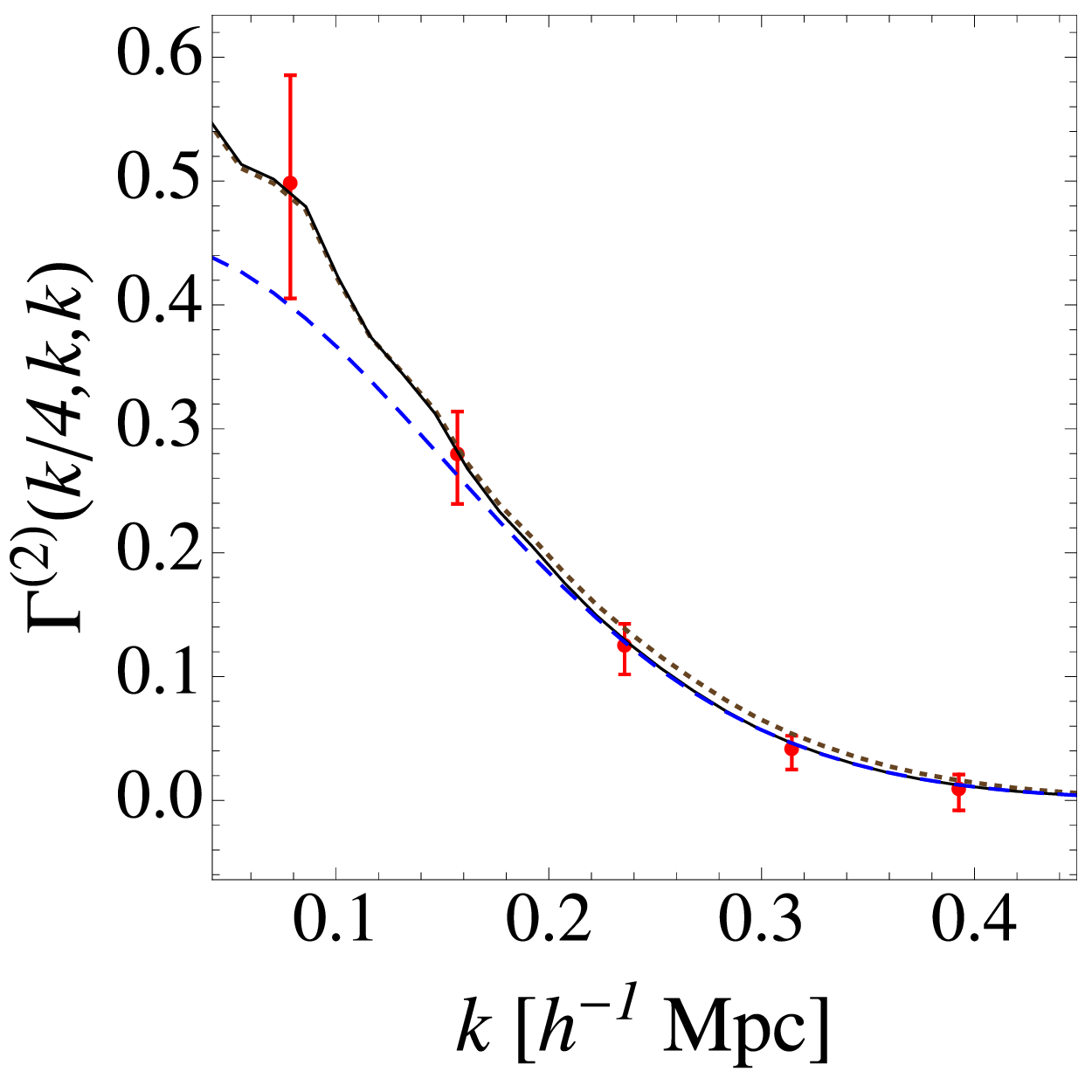,width=0.35\textwidth}\hspace{1cm}
\epsfig{file=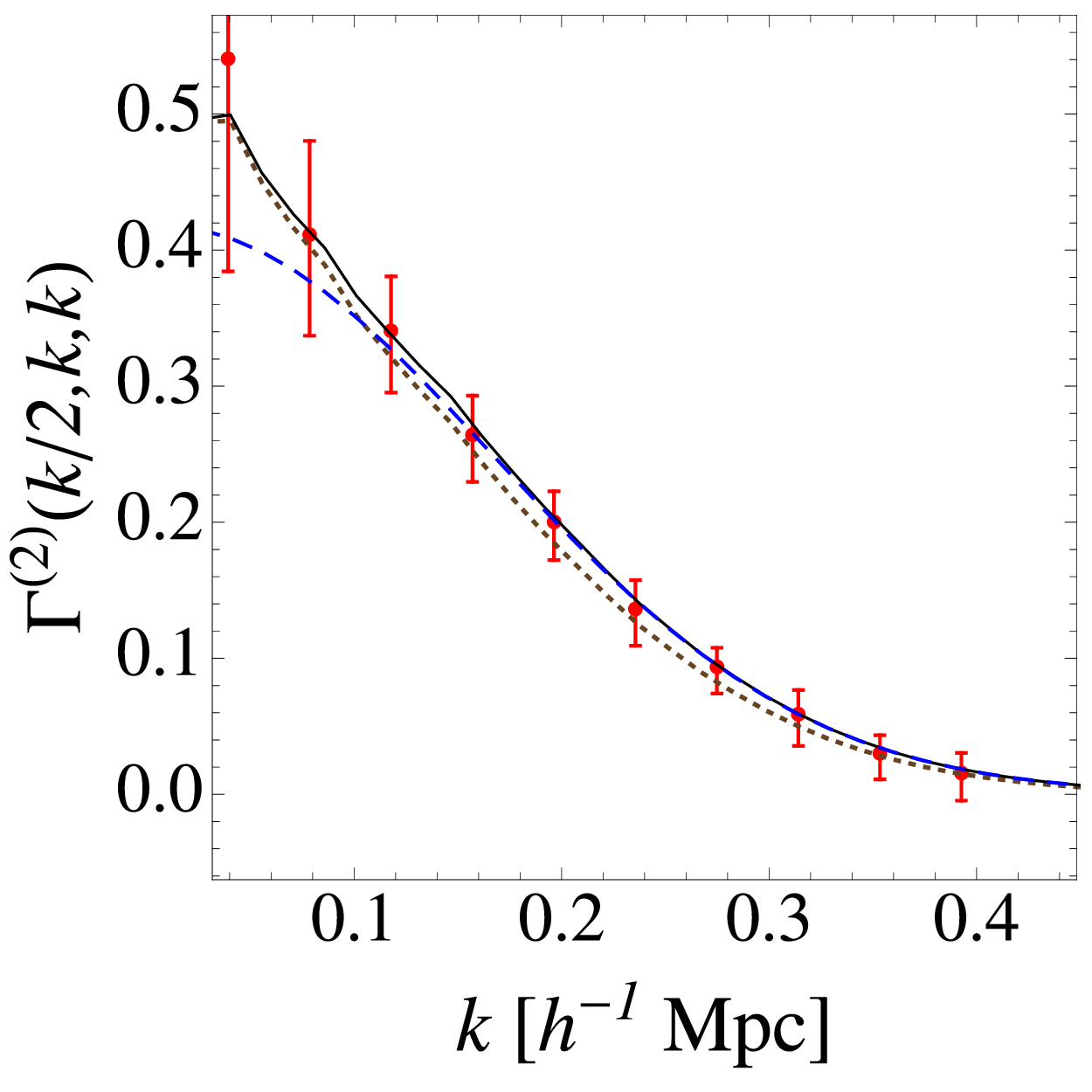,width=0.35\textwidth}
\epsfig{file=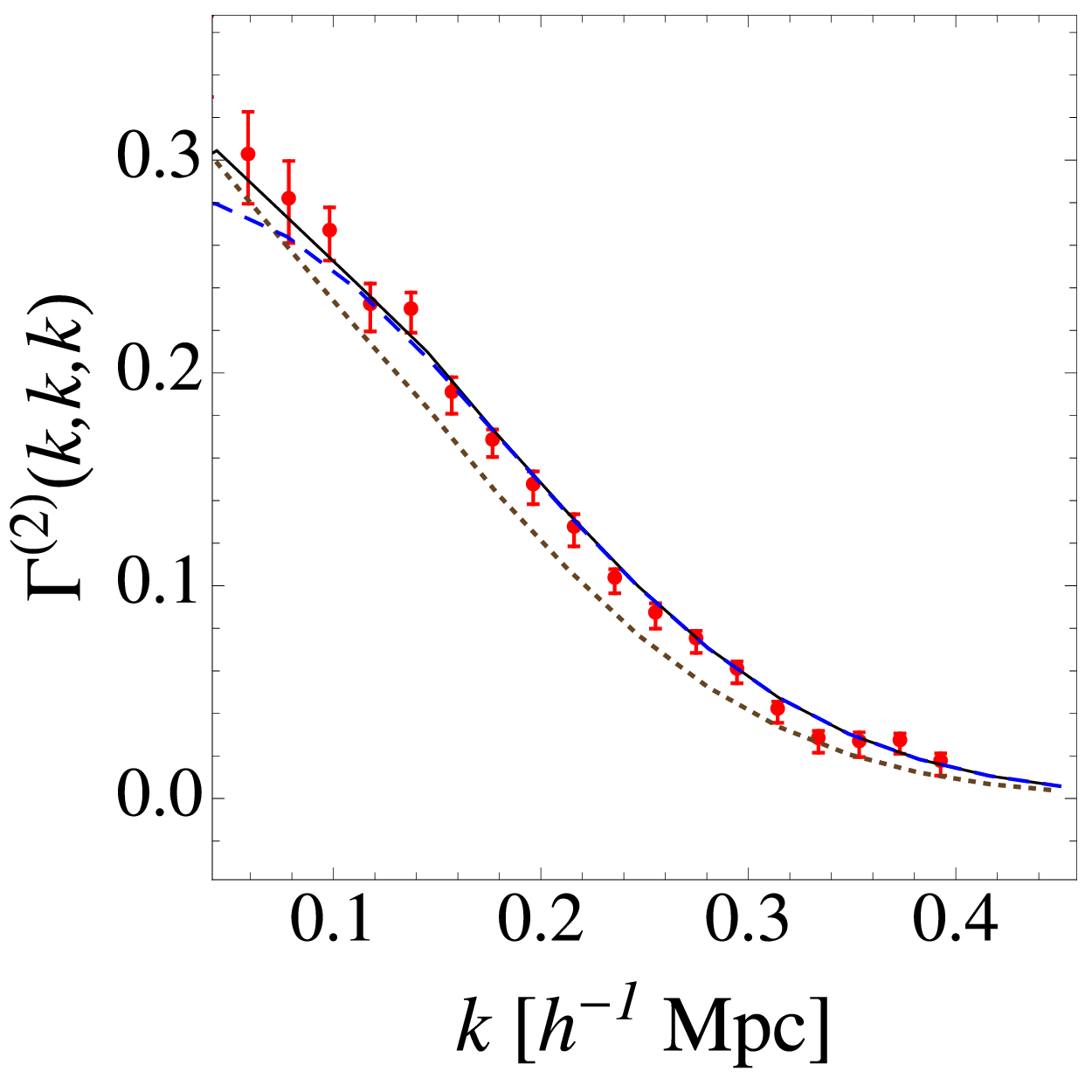,width=0.35\textwidth}\hspace{1cm}
\epsfig{file=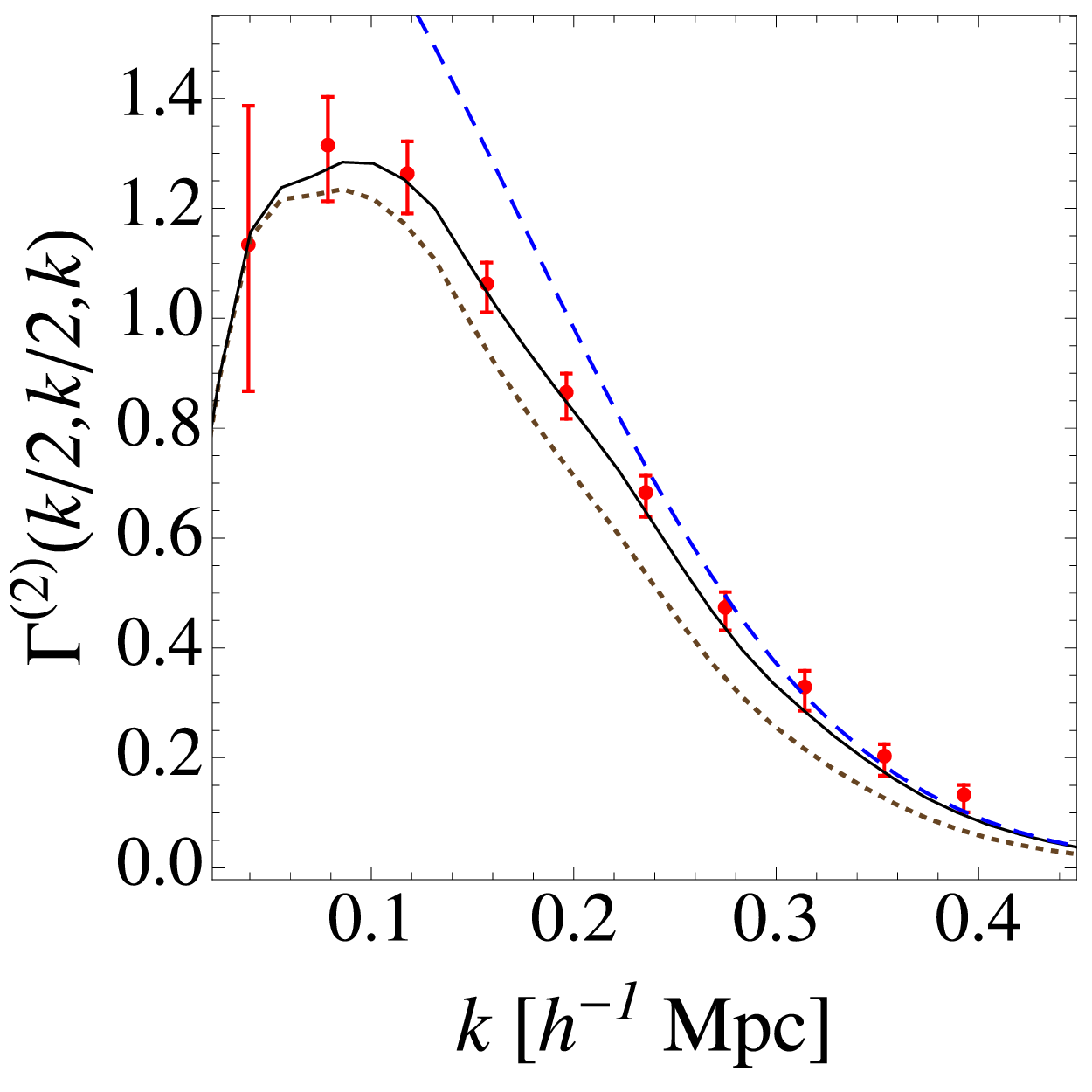,width=0.35\textwidth}
\caption{Comparison of the proposed interpolation scheme
  against  measurements of the two-point density propagator
in numerical simulations at $z=0$. Solid line is the prediction
  from Eq.~(\ref{regGammaponeloop}) including the only most growing mode into $\Gamma^{(2){\rm
    tree}}$ and $\delta \Gamma^{(2){\rm 1-loop}}$ after it is corrected for
binning as discussed in the text; the dashed line is when the binning is not taken into account and the dotted line
is the prediction when the one-loop correction is not taken into account.
Symbols with error bars are the corresponding measurements
in 50 runs of $L_{\lbox} = 1280 h^{-1}$ Mpc each. 
Different panels
corresponds to different wave-vector configurations (or ``triangle
shapes'') as detailed in the y-axis labels (and main text). The results are plotted as a function of $k=k_{3}$.
}
\label{regGamma2Num}
\end{center}
\end{figure}

As the prescription we are advocating here does not give significant
differences for the two-point propagator (already studied in \cite{2006PhRvD..73f3520C}) we focus our analysis in the
three-point propagator.

As shown in the previous paragraph the prescriptions in
Eqs.~(\ref{regGammaponeloop}, \ref{regGammap}) give non-trivial behaviors for the three-point
propagator. These prescriptions can be compared against
measurements in numerical simulations provided one can measure
cross-bispectra between initial conditions $\delta_0(\vk)$ and  the final density
fields $\delta(\vk,\eta)$. The estimator for the three-point propagator
was introduced in \cite{2008PhRvD..78j3521B};
\begin{equation}
\Gamma^{(2)}_{\delta}(k_1,k_2,k_3)=\frac{1}{2 P_0(k_1)P_0(k_2)}\frac{1}{N}\sum_{\vk_i \vk_j} \sum_{\vk_l}\delta(\vk_l,s) 
\times\delta_0(-\vk_i)\delta_0(-\vk_i), 
\label{algorithm}
\end{equation}
where the sum runs over Fourier modes $\vk_i$ in the $|\vk_1|$ bin,
$\vk_j$ in the  $|\vk_2|$ bin and $\vk_l$ in a bin $|\vk_3|$ such that
$|\vk_1-\vk_2| \le |\vk_3| \le \vk_1+ \vk_2$, and $N$ is the number of
terms in the triple sum. Equation (\ref{algorithm}) is valid for initial conditions set
in growing mode for which the only measurable quantity is the
contraction $\Gamma_{a b_1 b_2} u_{b_1} u_{b_2}$ with $u=(1,1)$ (and
$a=1=\delta$ in our case). In addition it assumes Gaussian
initial conditions, see \cite{2010PhRvD..82h3507B} otherwise.
We note that a similar expression holds for the velocity
divergence propagator (i.e. $a=2$), but its study is beyond the scope of
this paper.

We used Eq.~(\ref{algorithm}) to measure the three-point propagator in
a set of $50$ N-body simulations, each containing
$N_{par}=640^3$ particles within a comoving box-size of side $L_{box}=1280
h^{-1}{\rm Mpc}$. The total comoving volume of our simulations is
approximately $105~(h^{-1} {\rm Gpc})^3$. Cosmological parameters were
chosen as $\Omega_m=0.27$, $\Omega_{\Lambda}=0.73$,  $\Omega_b=0.046$
and $h=0.72$, together with scalar spectral index $n_s=1$ and normalization $\sigma_8=0.9$. The simulations were run using  Gadget2~\cite{2005MNRAS.364.1105S} with initial conditions set at $z_i=49$ using 2nd order Lagrangian Perturbation Theory (2LPT)~\cite{1998MNRAS.299.1097S,2006MNRAS.373..369C}.

Before comparing  theoretical predictions and numerical results it is
important 
to account for binning effects since correlations of modes within bins implicitly change
the shape of the predicted high-order propagators. Hence we will proceed by computing the predictions for binned modes. This is easily
accomplished as follows. Let us denote by an overline the bin average, e.g.
\begin{equation}
\overline\Gamma^{(2)}(k_{1},k_{2},k_{3})=\frac{1}{N_{\bin}\overline{P}(k_{1})\overline{P}(k_{2})}
\int_{\mB_{1}}\dd^3\vq_{1}
\int_{\mB_{2}}\dd^3\vq_{2}
\int_{\mB_{3}}\dd^3\vq_{3}\,
P(q_{1})\,P(q_{2})\,
\Gamma^{(2)}(q_{1},q_{2},q_{3})\Dirac(\vq_{3}-\vq_{1}-\vq_{2})
\end{equation}
where $\vq_{i}$ is in bin $\mB_{i}=\{k_{i}-\delta k/2,k_{i}+\delta
k/2\}$ and $N_{\bin}$ is the normalization 
\begin{equation}
N_{\bin}=\int_{\mB_{1}}\dd^3\vq_{1}
\int_{\mB_{2}}\dd^3\vq_{2}
\int_{\mB_{3}}\dd^3\vq_{3}\,
\Dirac(\vq_{3}-\vq_{1}-\vq_{2}),
\end{equation}
and 
\begin{equation}
\overline{P}(k_{i})=\int_{\mB_{i}}\dd^3\vq_{i} P(q_{i})/\int_{\mB_{i}}\dd^3\vq_{i}.
\end{equation}
Then writing the Dirac $\delta$-function as
$\Dirac(\vk)=\int\frac{\dd^3\vu}{(2\pi)^3}\exp(-\ii \vk.\vu)$ we finally have,
\begin{equation}
\overline{\Gamma}^{(2)}(k_{1},k_{2},k_{3})
=\frac{32\pi}{N_{\bin}}
\int_0^\infty u^2 d u
\int_{\mB_{1}}q_1^2\, j_{0}(q_{1}u) d q_{1}
\int_{\mB_{2}}q_2^2\, j_{0}(q_{2}u) d q_{2}
\int_{\mB_{3}}q_3^2\, j_{0}(q_{3}u) d q_{3}\
\Gamma^{(2)}(q_{1},q_{2},q_{3})
\end{equation}
with $N_{\bin}\approx 8 \pi^2 k_{1}k_{2}k_{3}(\delta k)^3$. This
expression, where $\Gamma^{(2)}$ is the model prediction from
Eq.~(\ref{regGammaponeloop}), is the one we use to compare with
measurements obtained in numerical simulations. Notice that we are  now
using three wave-modes as arguments with $\vk_3=\vk_1+\vk_2$ being the ``outgoing'' momenta.
More precisely, in our computations we choose the central bin value to be $k_{i}=4\,i \times 2\pi/L_{\lbox}$ and the bin 
width to be fixed and given by  $\delta k=4 \times 2\pi/L_{\lbox}$. Specific geometries, i.e. ratio of wave modes, are then defined from 
the central values of the bins. 

Those comparisons are shown in Fig. \ref{regGamma2Num} for four
different shapes, ``almost
squeezed'' ($k_{1} = k/4$ ; $k_{2} = k_{3} = k$),  ``elongated''
($k_{1} = k/2$ ; $k_{2} = k_{3} = k$), ``equilateral'' (where $k_{1} = k_{2} = k_{3}=k$) and  ``colinear'' ($k_{1} = k_{2} = k/2$ ; $k_{3} = k$). (Note that 
these configurations are not the same as in Fig. \ref{G2reg}. This is due to the fact that the ``anti-colinear'' and ``almost squeezed'' configurations presented there exhibited too large signal-to-noise ratios). Remarkably all measured configurations 
show a very good agreement  between the numerical results (points with error bars)
and the proposed prescription (solid lines). The dashed lines show the prediction before the binning corrections. The latter can
be quite large specifically at large scale (as expected) and furthermore in some configurations one-loop term corrections significantly improves
the predictions when compared to the numerical results.  One can also note that the size of the error bars change from configuration to configuration. This is due to the fact that, for a given $k_{3}$ the number of available modes in all three bins vary strongly.

Overall, these results indicate that the proposed interpolation scheme
works remarkably well when compared to measurements in simulations in an extended
$k$-range, which in turns is a very important step towards the
accurate modeling of polyspectra.

\section{Application: calculating the bispectrum}
\label{sec:bispectra}

\begin{figure}[ht]
\begin{center}
\epsfig{file= 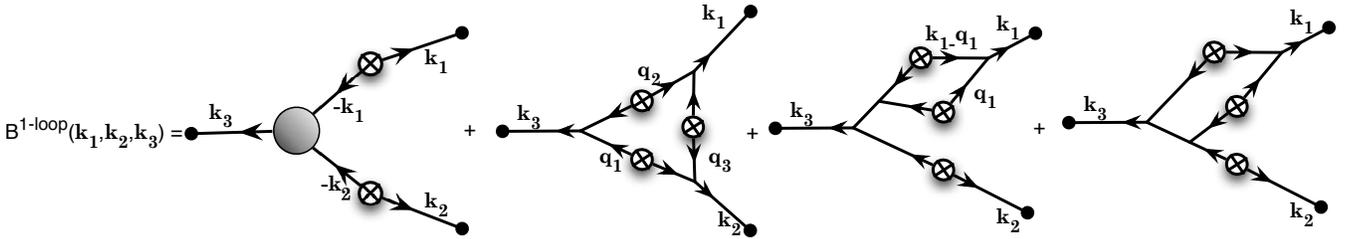,width=18cm}
\caption{The contributions to the bispectrum up to one loop order. In
  the first diagram the expressions of $\Gamma^{(2)}$ and
  $\Gamma^{(1)}$ should incorporate their own  one loop correction ;
  they are computed using formulae
    (\ref{regGammaponeloop}) at one loop order. In the other 3
  diagrams, the propagators correspond to their regularized tree order
  expressions;  they are computed using formulae (\ref{Gammaploops})
. There is thus no need at this order to include their one-loop correction, in particular to $\Gamma^{(3)}$.}
\label{BispectreContribs}
\end{center}
\end{figure}

We are interested in the computation of $B_{abc}(\vk_{1},\vk_{2},\vk_{3})$, the bispectrum of the components $a,b,c$ of the cosmic fluids,
\begin{equation}
\mg\Psi_{a}(\vk_{1},\eta)\Psi_{b}(\vk_{2},\eta)\Psi_{c}(\vk_{3},\eta)\md = \Dirac(\vk_{1}+\vk_{2}+\vk_{3})\,B_{abc}(\vk_{1},\vk_{2},\vk_{3},\eta)
\end{equation}
Such bispectra can be computed from a resummation of product of $\Gamma$ functions.
This is an extension of (\ref{B1loop}) and this is illustrated in Fig. \ref{BispectreContribs}. In particular if one wants to incorporate all one-loop 
effects one should include three types of diagrams:  the first one involves the one-loop correction of the propagators, the second, third
and fourth correspond to intrinsic one-loop contribution to the bispectra. 

More precisely, for growing mode initial conditions, the bispectrum takes the form,
\begin{eqnarray}
B_{abc}(\vk_{1},\vk_{2},\vk_{3},\eta)&=&
2\,\Gamma^{(2)\oneloop}_{a++}(\vk_{2},\vk_{3},\eta) G^{\oneloop}_{b+}(\vk_{2},\eta) G^{\oneloop}_{c+}(\vk_{3},\eta)P_{0}(k_{2})P_{0}(k_{3})
+ \sym \ {\rm (2\ terms)} \nonumber\\
&+&8\,\Dirac(\vk_{1}+\vq_{2}-\vq_{3})\,\Dirac(\vk_{2}+\vq_{3}-\vq_{1})\,\Dirac(\vk_{3}+\vq_{1}+\vq_{2})\,
P_{0}(q_{1})\,P_{0}(q_{2})\,P_{0}(q_{3})
\nonumber\\
&&\hspace{.5cm}\times\,
\Gamma^{(2)\,\tree}_{a++}(\vq_{3},-\vq_{2},\eta)\,
\Gamma^{(2)\,\tree}_{b++}(\vq_{1},-\vq_{3},\eta)\,
\Gamma^{(2)\,\tree}_{c++}(\vq_{2},-\vq_{1},\eta)+\sym\ {\rm (2\ terms)}\nonumber\\
&+&12\, 
\Gamma^{(3)\,\tree}_{a+++}(-\vk_{2}+\vq,-\vq,-\vk_{3},\eta)\,
\Gamma^{(2)\,\tree}_{b++}(\vk_{2}-\vq,\vq,\eta)\,
G^{\tree}_{c+}(\vk_{3},\eta)
\nonumber\\
&&\hspace{.5cm}\times\,P_{0}(\vert \vk_{2}-\vq\vert)\,P_{0}(q)\,P_{0}(k_{3})
+\sym\ {\rm (2\ terms)}
\end{eqnarray}
where 
all the propagators are computed using their regularized form, whether
 it is at tree order as in (\ref{Gammaploops}) or including one-loop correction as
in (\ref{regGammaponeloop}).
Integrals over the wavevectors $\vq_{1}$, $\vq_{2}$ and $\vq_{3}$ when present are implicit, and the symmetric terms are obtained
by a simultaneous permutation of the indices $abc$ and the wave modes $\vk_{1}$, $\vk_{2}$ and $\vk_{3}$. Note that in this formulation 
the $\Gamma^{(p)}$ functions are assumed to be symmetric functions of their arguments. Because of that the last 2 diagrams that appear in 
Fig. \ref{BispectreContribs} are automatically included.

For its practical evaluation the non trivial part is the one-loop expression of the $\Gamma_{a++}^{(2)}$ propagators. Its properties
were discussed in \cite{2008PhRvD..78j3521B}. We give in the appendix their actual values for both the density and the velocity
fields. They are the crucial ingredient to use to compute the bispectrum at one-loop order.

\begin{figure}[ht]
\begin{center}
\epsfig{file= 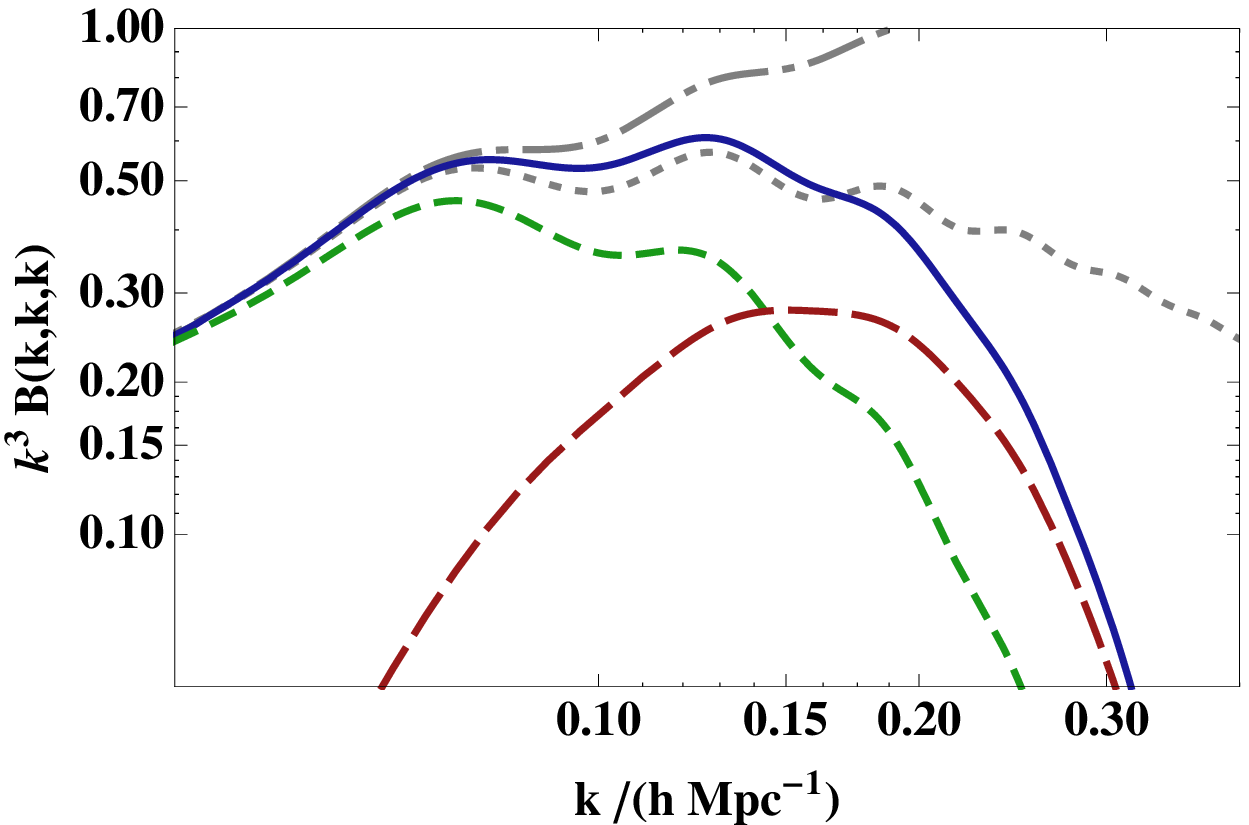,height=4cm}
\epsfig{file= 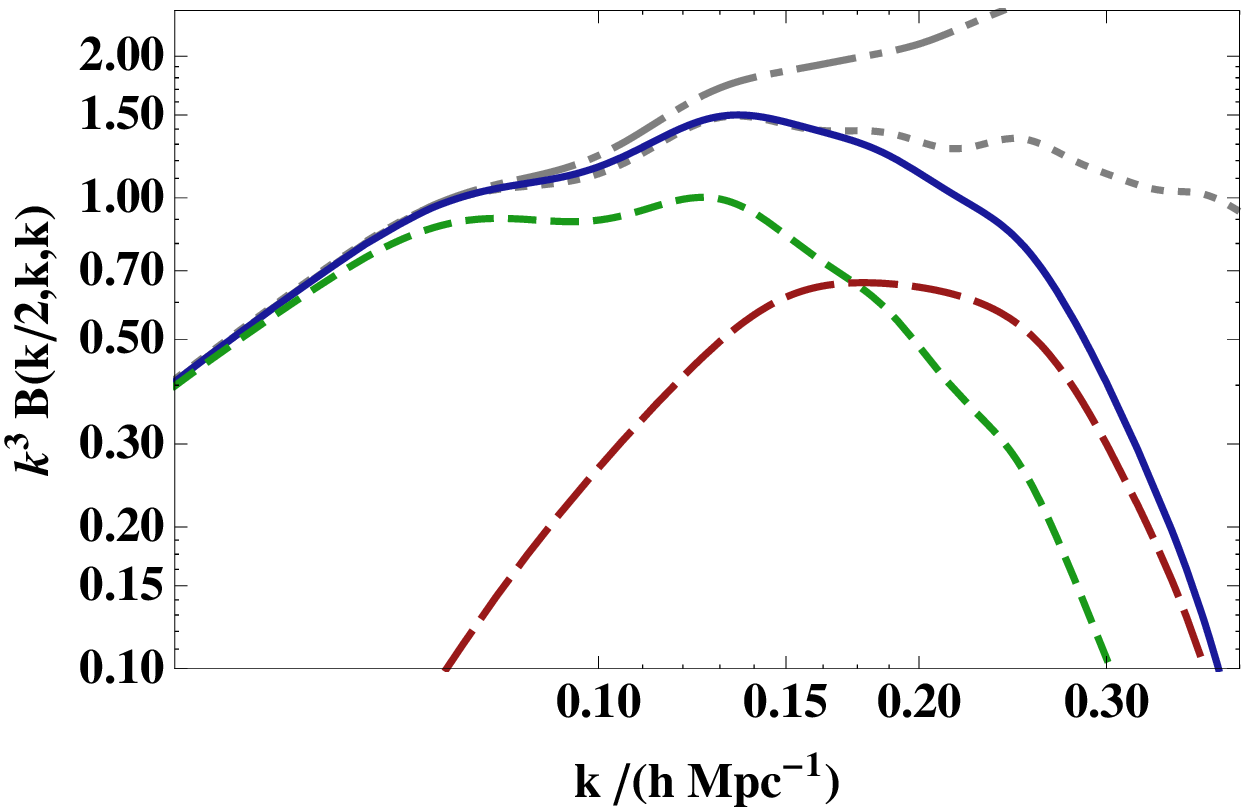,height=4cm}
\epsfig{file= 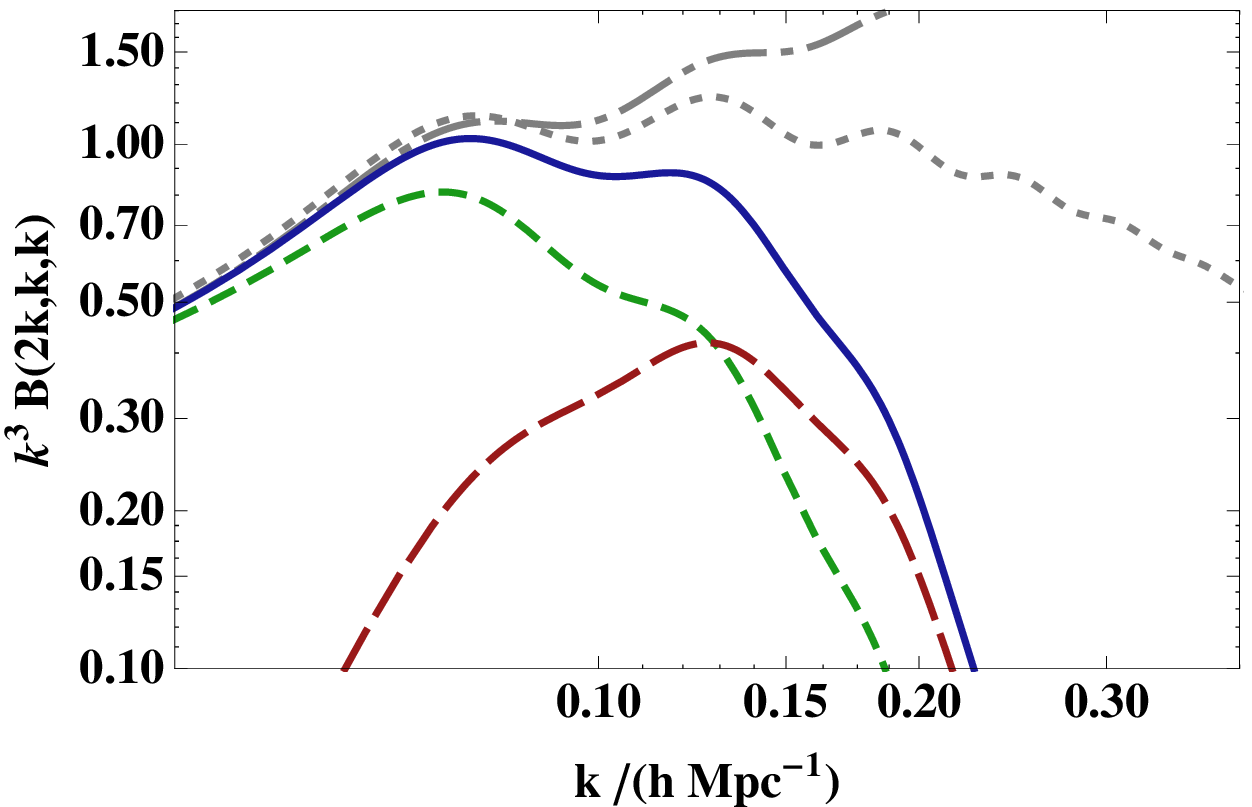,height=4cm}
\caption{The shape of the density bispectrum, $B_{111}(k_{1},k_{2},k_{3})$, as a function of scale for different configurations of the wavevectors.
The (gray) dotted line is the standard tree order prediction for the
bispectrum and the (gray) dot-dashed the standard
  one-loop result; the (green) short dashed line corresponds to the first contribution in the diagrammatic expansion of Fig. \ref{BispectreContribs}, the (red) long dashed line to the 2 others and the solid line the resulting reconstructed bispectrum. The first panel shows $k^{3}B(k,k,k)$, e.g. equilateral configuration. The second corresponds to a lightly squeezed configuration,  $k^{3}B(k/2,k,k)$, and the last to a flatten configuration, $k^{3}B(2k,k,k)$. The predictions are made for $z=0.5$.}
\label{BispectreBumps}
\end{center}
\end{figure}

\begin{figure}[ht]
\begin{center}
\epsfig{file=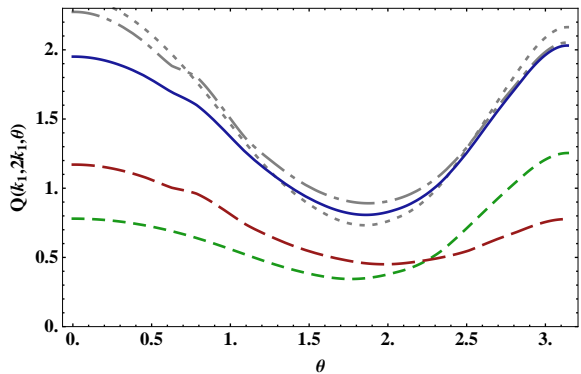,scale=.5}
\caption{The reduced bispectrum as a function of the relative angle
  between the wave-vectors $\theta$ for $k_{1}=0.1h$/Mpc and $k_{2}=0.2h$/Mpc. Lines follow the same convention as in Fig.~\ref{BispectreBumps}.}
\label{BispectreTheo}
\end{center}
\end{figure}

We present in Figs.~\ref{BispectreBumps} and \ref{BispectreTheo} the
resulting shape of the density bispectrum for a standard $\Lambda$CDM
model at $z=0.5$.  Figure \ref{BispectreBumps} shows the scale dependence of the bispectrum (multiplied by $k^3$ to make it less scale dependent). It makes clear that the contribution of the first diagram and that of the 3 others correspond to different scales, each producing one bump at different scales. This is reminiscent to what the RPT calculations give for the power spectrum~\cite{2006PhRvD..73f3519C}. 

In turn, Fig~\ref{BispectreTheo} explores the resulting angular dependence of the reduced bispectrum.
Here we plot
\begin{equation}
Q(k_{1},k_{2},k_{3})=\frac{1}{P_{0}(k_{1})P_{0}(k_{2})+P_{0}(k_{2})P_{0}(k_{3})+P_{0}(k_{3})P_{0}(k_{1})}B_{111}(\vk_{1},\vk_{2},\vk_{3})
\end{equation}
where the expression of the power spectrum is kept at linear
order. The result is expressed for fixed values of $k_{1}=0.1\,h$/Mpc
and $k_{2}=0.2\,h$/Mpc and 
as a function of their relative angle, $\theta$, so that
$k_{3}^2=k_{1}^2+k_{2}^2+2k_{1}k_{2}\cos\theta$. The plot compares the
naked ``tree'' order result (dotted line) with our prediction or the perturbative calculations
(same convention as in Fig. \ref{BispectreBumps}). 
Detailed comparisons of such predictions with $N$-body results is left for future studies.

\section{Conclusions}
\label{sec:conclusions}

In this paper we propose a systematic interpolation scheme that aims at describing the (multi-point) 
propagators in such a way that their expressions interpolate between their perturbation theory forms - to any loop order - 
and their large-$k$ behavior obtained from non-perturbative re-summations. 
This scheme is based on a separation of scales between the long wave-modes, whose effects are fully resumed and lead to 
the large-$k$ behavior, and  the short wave-modes that are fully taken into account in the perturbation theory treatment.
Our prescription is different from the exponentiation scheme proposed in \cite{2006PhRvD..73f3520C} but departs from it only very weakly.
Our new prescription however is very general and can be used for any loop order calculations and for any propagator. Furthermore, 
this construction is totally unambiguous. We note that it can even be used in the context of primordial non-Gaussianities
although new terms arise only at two loop order.

As the construction proposed here has a very general range of application, it should in principle be tested for a large variety of quantities, 
from two-point and multi-point propagators of the density field to the ones for the velocity field. We proposed here some comparisons with 
N-body results for the quantities that are of most interest  for the use of the multi-point propagators in the context of the $\Gamma$-expansion 
applied to the calculation of the  density power spectrum. Because of the property we mentioned in the previous paragraph, this prescription 
for the two-point propagator is found to give very accurate results for the two-point propagator. We leave for further studies the impact of
two-loop effects. In this work we further check the validity of our prescription against numerical results for the one-loop level 
of the density three-point propagator. These comparisons are presented in Fig. \ref{regGamma2Num}. 
We found that it gives a satisfactory form for a large range of configurations, i.e. interpolating the low-$k$ one-loop result
with the high-$k$ exponential decay, even when the signs of tree order and one-loop forms differ. 

In the context of the $\Gamma$-expansion approach this construction therefore provides us with the necessary recipes for constructing 
poly-spectra incorporating any order of perturbation theory results. In coming papers we will explicitly compare predictions of the $\Gamma$-expansion 
for the power spectra with numerical simulations when higher order PT loop corrections are included. 
These prescriptions provide a good opportunity for giving the explicit form of the bispectra in the context of the $\Gamma$-expansion. The 
explicit mathematical forms are given in Section \ref{sec:bispectra} when bispectra are computed up to one-loop order. Such expressions
make use in particular of the three-point propagators at one-loop order whose explicit forms are given in the appendix.
We found in particular that the bispectra terms can be separated in a tree order contribution and coupling terms contributions
that contribute at different scales, i.e. in subsequent bumps,  in a similar way to the power spectrum.
Comparison of the proposed forms for the bispectrum with $N$-body results is however left for further studies. 

We finally note that such a construction is restricted here to the case of a single pressure-less fluid. Whether it could be used in cases the fluid content of the universe is richer (with extra degrees of freedom carried by baryons, see \cite{2010PhRvD..81b3524S,2011arXiv1109.3400B}, massive neutrinos, dark energy components, see \cite{2011JCAP...06..019B,2011JCAP...03..047S}, etc.) is still largely an open issue. Results
obtained in  \cite{2011arXiv1109.3400B} however suggest that the effects of {\sl adiabatic} long wavelength modes can be resummed and therefore could be incorporated in a large variety of cases.

\begin{acknowledgements}

We thank the Programme National de Cosmologie et Galaxies for their support. 
M.C. acknowledges support by the Spanish Ministerio de Ciencia e Innovacion (MICINN), project AYA2009-13936, Consolider-Ingenio CSD2007- 00060, European Commission Marie Curie Initial Training Network CosmoComp (PITN-GA-2009-238356), research project 2009-SGR-1398 from Generalitat de Catalunya and the Juan de la Cierva MICINN program. R.S. was partially supported by grants NSF AST-1109432 and NASA NNA10A171G.

\end{acknowledgements}

\bibliography{RePeaT,masterbiblio}

\appendix
\section{Explicit expression of the one-loop $\Gamma^{(2)}$ propagators}
\newcommand{\LFonc}{{\cal L}}
\newcommand{\LFoncW}{{\cal W}}

We give here the explicit expressions of the late time component (most growing
term) of the 1-loop expression of the three-point propagator
(for growing mode initial conditions). The final result is obtained by the explicit integration over the wave mode
$q$ depending on the linear power spectrum $P_{0}(q)$,
\begin{equation}
\Gamma^{(2)\oneloop}_{1++}(k_{1},k_{2},k_{3})=4\pi e^{4\eta}\int q^2\dd q \Gamma^{(2)\oneloop}_{1++}(k_{1},k_{2},k_{3},q)\,P_{0}(q)
\end{equation}
where the result is expressed in terms of the 3 norms, $k_{1}$, $k_{2}$, $k_{3}$ so that $k_{3}^2=k_{1}^2+k_{2}^2+2\vk_{1}.\vk_{2}$. Note that 
these three norms obey the triangular inequality.

Note that $\Gamma^{(2)\oneloop}_{i++}(k_{1},k_{2},k_{3},q)$  obeys the following asymptotic behaviors,
\begin{eqnarray}
\Gamma^{(2)\oneloop}_{i++}(k_{1},k_{2},k_{3},q) &\to& -\frac{k_{3}^2}{6q^2}\Gamma^{(2)\tree}_{i++}(k_{1},k_{2},k_{3}),
\end{eqnarray}
and $q\to 0$.
We also have
\begin{eqnarray}
\Gamma^{(2)\oneloop}_{1++}(\vk_{1},\vk_{2})&\to& \frac{\vk_{1}.\vk_{2}}{2 k_{1}^2}f(k_{2}),\\
\Gamma^{(2)\oneloop}_{2++}(\vk_{1},\vk_{2})&\to& \frac{\vk_{1}.\vk_{2}}{2 k_{1}^2}g(k_{2}),
\end{eqnarray}
when $\vk_{1}\to 0$ and where the functions $f(k)$ and $g(k)$ are defined in Eqs. (\ref{functionsofk}). These 2 asymptotic regimes can be used as internal checks.

The function $\Gamma^{(2)\oneloop}_{1++}(k_{1},k_{2},k_{3},q)$ has been obtained after integration of the angular variable in the loop expressions.
The result can be expressed in terms of the functions
\begin{eqnarray}
\LFonc \left (k_1 \right) &=&\log\frac{(k_{1}+q)^2}{(k_{1}-q)^2}\\
\LFoncW\left (k_1, k_2, k_3 \right)&=&\log \left(\frac{-4 k_3^2 q^2-2 \left(k_1^2-q^2\right) \left(k_2^2-q^2\right)-4 k_3 q \sqrt{\left(-k_1^2-k_2^2+k_3^2\right) q^2+k_1^2 k_2^2+q^4}}{-4 k_3^2
   q^2-2 \left(k_1^2-q^2\right) \left(k_2^2-q^2\right)+4 k_3 q \sqrt{\left(-k_1^2-k_2^2+k_3^2\right) q^2+k_1^2 k_2^2+q^4}}\right)
\end{eqnarray}
and it reads,
\begin{eqnarray}
&&\Gamma^{(2)\oneloop}_{1++}(k_{1},k_{2},k_{3},q)=\nonumber\\
&&\frac {k_2 \left (k_1^2 - q^2 \right)^2 \left (7 k_3^2 q^2 + 2 k_3^4 - 
      9 q^4 \right)} {29568\ k_1^2 q^5 \sqrt {q^2 \left (k_2^2 -    k_3^2 + q^2 \right) + k_1^2 \left (k_3^2 - q^2 \right)}}\ \LFoncW\left (k_1, k_3, 
    k_2 \right)
        +\sym{\rm\ (k_{1}\leftrightarrow k_{2})}
  \nonumber\\
&&        +
\frac {3 k_3^3 \left (k_1^2 - q^2 \right)^2 \left (k_2^2 -q^2 \right)^2} {17248\ k_1^2 k_2^2 q^5 \sqrt {q^2 \left (-k_2^2 + k_3^2 + q^2 \right) + k_1^2 \left (k_2^2 - q^2 \right)}}\ \LFoncW \left (k_1, k_2,  k_3 \right)
\nonumber\\
&&-\frac{\left (k_1^2 - 
     q^2 \right)^2} {1655808\ k_1^9 k_2^2 q^5}\ 
     \left[165 \left (k_2^2 - 
        k_3^2 \right)^3 q^6 -   k_1^{10} \left (582 k_2^2 + 791 k_3^2 + 1947 q^2 \right) + \right.\nonumber\\&& \hspace{.5cm}\left.
  +  k_1^6 \left (-1113 k_3^2 q^4 + 179 k_3^4 q^2 - 
        3 k_2^4 \left (211 k_3^2 + 2155 q^2 \right) + 
        k_2^2 \left (758 k_3^2 q^2 + 390 k_3^4 - 
           7284 q^4 \right) + 492 k_2^6 + 151 k_3^6 + 
        219 q^6 \right)                                                                  \right.\nonumber\\&& \hspace{.5cm}\left.+
           k_1^8 \left (1787 k_3^2 q^2 + 
        k_2^2 \left (2120 k_3^2 + 6891 q^2 \right) + 51 k_2^4 +  
        601 k_3^4 + 1689 q^4 \right)                                        \right.\nonumber\\&& \hspace{.5cm}\left.+
             k_1^4 q^2 \left (405 k_3^2 q^4 + 
        k_2^4 \left (7269 q^2 - 1677 k_3^2 \right) - 
        321 k_3^4 q^2 +      k_2^2 \left (-4376 k_3^2 q^2 + 391 k_3^4 + 
           471 q^4 \right) + 1017 k_2^6 + 269 k_3^6 \right)     \right.\nonumber\\&& \hspace{.5cm}\left.-
     3 \left (k_2^2 - k_3^2 \right) k_1^2 q^4 \left (k_2^2 \left (285 q^2 -473 k_3^2 \right) - 
        17 \left (9 k_3^2 q^2 + 5 k_3^4 \right) + 
        558 k_2^4 \right) + 39 k_1^{12} \right]\         \LFonc \left (k_1 \right) 
+\sym{\rm\ (k_{1}\leftrightarrow k_{2})}
  \nonumber\\
&&
+
\frac {\left (7 k_3^2 q^2 + 2 k_3^4 - 
      9 q^4 \right)} {236544\ k_1^2 k_2^2 k_3^5 q^5}
       \left[
-12 \left(k_1^2-k_2^2\right)^2 q^6-3 \left(k_1^2-k_2^2\right)^2 \left(k_1^2+k_2^2\right) q^4
\right.\nonumber\\&& \hspace{.5cm}\left.
 -2 k_3^{10}-15 \left(k_1^2+k_2^2\right) k_3^8+k_3^6 \left(46 \left(k_1^2+k_2^2\right) q^2+16
   \left(k_1^2-k_2^2\right)^2-2 q^4\right)
   \right.\nonumber\\&& \hspace{.5cm}\left.
   +k_3^4 \left(-47 \left(k_1^2+k_2^2\right) q^4-8 \left(7 k_1^4-17 k_2^2 k_1^2+7 k_2^4\right)
   q^2+\left(k_1^2+k_2^2\right) \left(k_1^4-10 k_2^2 k_1^2+k_2^4\right)+4 q^6\right)
   \right.\nonumber\\&& \hspace{.5cm}\left.
   +k_3^2 \left(8 \left(k_1^2+k_2^2\right) q^6+\left(52 k_1^4-96 k_2^2
   k_1^2+52 k_2^4\right) q^4+2 \left(k_1^2-k_2^2\right)^2 \left(k_1^2+k_2^2\right) q^2\right)  
        \right]
\LFonc\left (k_3 \right)       \nonumber \\ 
       &&+
\frac {1}{{6209280\ q^4 k_1^8 k_2^8 k_3^4}}
\left\{
420 k_1^6 k_2^6 k_3^{12}                                 
 \right.\nonumber\\&& \hspace{.5cm}\left.
 +\left[-2475 \left(k_1^6+k_2^6\right) q^8+300 k_1^2 k_2^2 \left(k_1^4+k_2^4\right) q^6+9090 k_1^4 k_2^4 \left(k_1^2+k_2^2\right)q^4
             \right.\right.\nonumber\\&& \hspace{1cm}\left.\left.
  -68910 k_1^6 k_2^6 q^2+5415 k_1^6 k_2^6 \left(k_1^2+k_2^2\right)\right] k_3^{10}
    \right.\nonumber\\&& \hspace{.5cm}\left.
+\left[135 \left(55 k_1^8-51 k_2^2 k_1^6-51 k_2^6 k_1^2+55
   k_2^8\right) q^8-45 k_1^2 k_2^2 \left(k_1^2+k_2^2\right) \left(663 k_1^4-811 k_2^2 k_1^2+663 k_2^4\right) q^6
       \right.\right.\nonumber\\&& \hspace{1cm}\left.\left.
   +\left(38925 k_2^4 k_1^8+9532 k_2^6
   k_1^6+38925 k_2^8 k_1^4\right) q^4-120370 k_1^6 k_2^6 \left(k_1^2+k_2^2\right) q^2
          \right.\right.\nonumber\\&& \hspace{1cm}\left.\left.
+15 k_1^6 k_2^6 \left(377 k_1^4+940 k_2^2 k_1^2+377 k_2^4\right)\right]
   k_3^8
       \right.\nonumber\\&& \hspace{.5cm}\left.
+\left[-135 \left(k_1^2+k_2^2\right) \left(55 k_1^8-201 k_2^2 k_1^6+156 k_2^4 k_1^4-201 k_2^6 k_1^2+55 k_2^8\right) q^8
       \right.\right.\nonumber\\&& \hspace{1cm}\left.\left.
+30 k_1^2 k_2^2 \left(1959
   k_1^8-3283 k_2^2 k_1^6-1907 k_2^4 k_1^4-3283 k_2^6 k_1^2+1959 k_2^8\right) q^6
          \right.\right.\nonumber\\&& \hspace{1cm}\left.\left.
-2 k_1^4 k_2^4 \left(k_1^2+k_2^2\right) \left(53220 k_1^4-178337 k_2^2
   k_1^2+53220 k_2^4\right) q^4
          \right.\right.\nonumber\\&& \hspace{1cm}\left.\left.
+10 k_1^6 k_2^6 \left(22261 k_1^4-45566 k_2^2 k_1^2+22261 k_2^4\right) q^2-15 k_1^6 k_2^6 \left(k_1^2+k_2^2\right) \left(805
   k_1^4-2306 k_2^2 k_1^2+805 k_2^4\right)\right] k_3^6
       \right.\nonumber\\&& \hspace{.5cm}\left.
+\left[45 \left(55 k_1^{12}-285 k_2^2 k_1^{10}+157 k_2^4 k_1^8+230 k_2^6 k_1^6+157 k_2^8 k_1^4-285
   k_2^{10} k_1^2+55 k_2^{12}\right) q^8
          \right.\right.\nonumber\\&& \hspace{1cm}\left.\left.
-15 k_1^2 k_2^2 \left(k_1^2+k_2^2\right) \left(1949 k_1^8-10643 k_2^2 k_1^6+20573 k_2^4 k_1^4-10643 k_2^6 k_1^2+1949
   k_2^8\right) q^6
          \right.\right.\nonumber\\&& \hspace{1cm}\left.\left.
+3 k_1^4 k_2^4 \left(19475 k_1^8-139742 k_2^2 k_1^6+246974 k_2^4 k_1^4-139742 k_2^6 k_1^2+19475 k_2^8\right) q^4
       \right.\right.\nonumber\\&& \hspace{1cm}\left.\left.
-30 k_1^6 k_2^6
   \left(k_1^2+k_2^2\right) \left(911 k_1^4-1570 k_2^2 k_1^2+911 k_2^4\right) q^2+45 k_1^6 k_2^6 \left(k_1^2-k_2^2\right){}^2 \left(13 k_1^4-4 k_2^2
   k_1^2+13 k_2^4\right)\right] k_3^4
          \right.\nonumber\\&& \hspace{.5cm}\left.
+\left[7560 k_1^6 k_2^6 \left(k_1^2+k_2^2\right) q^8+1260 k_1^6 k_2^6 \left(43 k_1^4-80 k_2^2 k_1^2+43 k_2^4\right)
   q^6+3150 k_1^6 k_2^6 \left(k_1^2-k_2^2\right)^2 \left(k_1^2+k_2^2\right) q^4\right] k_3^2
          \right.\nonumber\\&& \hspace{.5cm}\left.
-11340 q^8 k_1^6 k_2^6 \left(k_1^2-k_2^2\right)^2-2835 q^6
   k_1^6 k_2^6 \left(k_1^2-k_2^2\right)^2 \left(k_1^2+k_2^2\right)
\right\}.
\end{eqnarray}
It can be noted that the final expression is symmetric in $k_{1}$ and $k_{2}$.
The velocity component can be similarly constructed,

\begin{eqnarray}
&&\Gamma^{(2)\oneloop}_{2++}(k_{1},k_{2},k_{3},q)=\nonumber\\
&&\frac {k_2 \left (k_1^2 - q^2 \right)^2 \left(-5 k_3^2 q^2+8 k_3^4-3 q^4\right)} 
{29568\ k_1^2 q^5 \sqrt {q^2 \left (k_2^2 -    k_3^2 + q^2 \right) + k_1^2 \left (k_3^2 - q^2 \right)}}\ \LFoncW\left (k_1, k_3, 
    k_2 \right)
        +\sym{\rm\ (k_{1}\leftrightarrow k_{2})}
  \nonumber\\
&&        +
\frac {3 k_3^3 \left (k_1^2 - q^2 \right)^2 \left (k_2^2 -q^2 \right)^2} {4312\ k_1^2 k_2^2 q^5 \sqrt {q^2 \left (-k_2^2 + k_3^2 + q^2 \right) + k_1^2 \left (k_2^2 - q^2 \right)}}\ \LFoncW \left (k_1, k_2,  k_3 \right)
\nonumber\\
&&-\frac{\left (k_1^2 - 
     q^2 \right)^2} {1655808\ k_1^9 k_2^2 q^5}\ 
     \left[
     55 \left(k_2^2-k_3^2\right){}^3 q^6-k_1^{10} \left(194 k_2^2+1921 k_3^2+649 q^2\right)  +13 k_1^{12}
     \right.\nonumber\\&& \hspace{.5cm}\left.
      +k_1^6 \left(25 k_3^2 q^4-1759 k_3^4 q^2+k_2^4 \left(801 k_3^2-2155 q^2\right)+k_2^2 \left(-3414 k_3^2 q^2+306 k_3^4-2428 q^4\right)+164
   k_2^6+329 k_3^6+73 q^6\right)
     \right.\nonumber\\&& \hspace{.5cm}\left.   
     +k_1^8 \left(2913 k_3^2 q^2+k_2^2 \left(1176 k_3^2+2297 q^2\right)+17
   k_2^4+1579 k_3^4+563 q^4\right)
   \right.\nonumber\\&& \hspace{.5cm}\left.
   +k_1^4 q^2 \left(135 k_3^2 q^4+k_2^4 \left(321 k_3^2+2423 q^2\right)+333 k_3^4 q^2+k_2^2 \left(-784 k_3^2 q^2-1307
   k_3^4+157 q^4\right)+339 k_2^6+647 k_3^6\right)
   \right.\nonumber\\&& \hspace{.5cm}\left.
   -3 \left(k_2^2-k_3^2\right) k_1^2 q^4 \left(k_2^2 \left(121 k_3^2+95 q^2\right)-51 k_3^2 q^2+186 k_2^4-307
   k_3^4\right)  \right]\         \LFonc \left (k_1 \right) 
+\sym{\rm\ (k_{1}\leftrightarrow k_{2})}
  \nonumber\\
&&
+
\frac {\left (-5 k_3^2 q^2 + 8 k_3^4 - 3 q^4 \right)} {236544\ k_1^2 k_2^2 k_3^5 q^5}
       \left[
-12 \left(k_1^2-k_2^2\right)^2 q^6-3 \left(k_1^2-k_2^2\right)^2 \left(k_1^2+k_2^2\right) q^4
\right.\nonumber\\&& \hspace{.5cm}\left.
 -2 k_3^{10}-15 \left(k_1^2+k_2^2\right) k_3^8+k_3^6 \left(46 \left(k_1^2+k_2^2\right) q^2+16
   \left(k_1^2-k_2^2\right)^2-2 q^4\right)
   \right.\nonumber\\&& \hspace{.5cm}\left.
   +k_3^4 \left(-47 \left(k_1^2+k_2^2\right) q^4-8 \left(7 k_1^4-17 k_2^2 k_1^2+7 k_2^4\right)
   q^2+\left(k_1^2+k_2^2\right) \left(k_1^4-10 k_2^2 k_1^2+k_2^4\right)+4 q^6\right)
   \right.\nonumber\\&& \hspace{.5cm}\left.
   +k_3^2 \left(8 \left(k_1^2+k_2^2\right) q^6+\left(52 k_1^4-96 k_2^2
   k_1^2+52 k_2^4\right) q^4+2 \left(k_1^2-k_2^2\right)^2 \left(k_1^2+k_2^2\right) q^2\right)  
        \right]
\LFonc\left (k_3 \right)       \nonumber \\ 
       &&+
\frac {1}{{6209280\ q^4 k_1^8 k_2^8 k_3^4}}
\left\{
1680 k_1^6 k_2^6 k_3^{12}
 \right.\nonumber\\&& \hspace{.5cm}\left.
+\left(-825 \left(k_1^6+k_2^6\right) q^8-12440 k_1^2 k_2^2 \left(k_1^4+k_2^4\right) q^6+32290 k_1^4 k_2^4 \left(k_1^2+k_2^2\right)
   q^4           
    \right.\right.\nonumber\\&& \hspace{1cm}\left.\left.
-133850 k_1^6 k_2^6 q^2+17535 k_1^6 k_2^6 \left(k_1^2+k_2^2\right)\right) k_3^{10}
       \right.\nonumber\\&& \hspace{.5cm}\left.
+\left(45 \left(55 k_1^8-51 k_2^2 k_1^6-51 k_2^6 k_1^2+55
   k_2^8\right) q^8+15 k_1^2 k_2^2 \left(k_1^2+k_2^2\right) \left(1009 k_1^4-421 k_2^2 k_1^2+1009 k_2^4\right) q^6
          \right.\right.\nonumber\\&& \hspace{1cm}\left.\left.
-k_1^4 k_2^4 \left(50385 k_1^4+92816 k_2^2
   k_1^2+50385 k_2^4\right) q^4-81190 k_1^6 k_2^6 \left(k_1^2+k_2^2\right) q^2
             \right.\right.\nonumber\\&& \hspace{1cm}\left.\left.
+15 k_1^6 k_2^6 \left(683 k_1^4+1252 k_2^2 k_1^2+683 k_2^4\right)\right)
   k_3^8
          \right.\nonumber\\&& \hspace{.5cm}\left.
+\left(-45 \left(k_1^2+k_2^2\right) \left(55 k_1^8-201 k_2^2 k_1^6+156 k_2^4 k_1^4-201 k_2^6 k_1^2+55 k_2^8\right) q^8
       \right.\right.\nonumber\\&& \hspace{1cm}\left.\left.
+30 k_1^2 k_2^2 \left(235
   k_1^8-757 k_2^2 k_1^6-137 k_2^4 k_1^4-757 k_2^6 k_1^2+235 k_2^8\right) q^6
             \right.\right.\nonumber\\&& \hspace{1cm}\left.\left.
-4 k_1^4 k_2^4 \left(k_1^2+k_2^2\right) \left(345 k_1^4+628 k_2^2 k_1^2+345
   k_2^4\right) q^4
             \right.\right.\nonumber\\&& \hspace{1cm}\left.\left.
+10 k_1^6 k_2^6 \left(24507 k_1^4-50506 k_2^2 k_1^2+24507 k_2^4\right) q^2-15 k_1^6 k_2^6 \left(k_1^2+k_2^2\right) \left(1977 k_1^4-4010
   k_2^2 k_1^2+1977 k_2^4\right)\right) k_3^6
          \right.\nonumber\\&& \hspace{.5cm}\left.
+\left(15 \left(55 k_1^{12}-285 k_2^2 k_1^{10}+157 k_2^4 k_1^8+230 k_2^6 k_1^6+157 k_2^8 k_1^4-285 k_2^{10}
   k_1^2+55 k_2^{12}\right) q^8
             \right.\right.\nonumber\\&& \hspace{1cm}\left.\left.
-5 k_1^2 k_2^2 \left(k_1^2+k_2^2\right) \left(1949 k_1^8-10643 k_2^2 k_1^6+19341 k_2^4 k_1^4-10643 k_2^6 k_1^2+1949
   k_2^8\right) q^6
            \right.\right.\nonumber\\&& \hspace{1cm}\left.\left. 
  +k_1^4 k_2^4 \left(19475 k_1^8-93542 k_2^2 k_1^6+160734 k_2^4 k_1^4-93542 k_2^6 k_1^2+19475 k_2^8\right) q^4
            \right.\right.\nonumber\\&& \hspace{1cm}\left.\left.
-90 k_1^6 k_2^6
   \left(k_1^2+k_2^2\right) \left(67 k_1^4-106 k_2^2 k_1^2+67 k_2^4\right) q^2+15 k_1^6 k_2^6 \left(k_1^2-k_2^2\right){}^2 \left(13 k_1^4-4 k_2^2 k_1^2+13
   k_2^4\right)\right) k_3^4
             \right.\nonumber\\&& \hspace{.5cm}\left.
+\left(2520 k_1^6 k_2^6 \left(k_1^2+k_2^2\right) q^8+1260 k_1^6 k_2^6 \left(7 k_1^4-12 k_2^2 k_1^2+7 k_2^4\right) q^6-1260 k_1^6
   k_2^6 \left(k_1^2-k_2^2\right){}^2 \left(k_1^2+k_2^2\right) q^4\right) k_3^2
             \right.\nonumber\\&& \hspace{.5cm}\left.
-3780 q^8 k_1^6 k_2^6 \left(k_1^2-k_2^2\right){}^2-945 q^6 k_1^6 k_2^6
   \left(k_1^2-k_2^2\right){}^2 \left(k_1^2+k_2^2\right)
\right\}.
\end{eqnarray}

\end{document}